\documentclass[a4paper]{article}
\usepackage[authoryear,longnamesfirst]{natbib}
\usepackage{booktabs}
\usepackage{float}
\usepackage{dsfont}
\usepackage{placeins}
\usepackage[labelfont = bf]{caption}
\usepackage{graphicx}
\usepackage{enumerate}
\usepackage{subcaption}
\usepackage{orcidlink}
\usepackage{amsmath,amsfonts,amssymb}
\usepackage{geometry}
\begin{document}
\let\WriteBookmarks\relax
\def\floatpagepagefraction{1}
\def\textpagefraction{.001}

\title{Practical Guidelines for Goodness-of-Fit Testing for Right-Censored Data}
\author{Arnau G\'omez-Catal\`a, \orcidlinki{Guadalupe Gómez Melis}{0000-0003-4252-4884}, \orcidlinki{Klaus Langohr}{0000-0001-7075-9192} \\
        Department of Statistics and Operations Research,\\ Universitat Polit\`ecnica de Catalunya$-$BarcelonaTech, Barcelona, Spain}
\date{October 27, 2026}


%

\maketitle

\begin{abstract}
In survival studies, right censoring often prevents the observation of subjects' complete outcomes, complicating inference about the underlying distribution. Evaluating whether a given distribution adequately fits right-censored data is therefore essential. However, guidance on the relative performance of available goodness-of-fit tests under different data conditions remains limited.

In this work, we present a comparative assessment of three classical goodness-of-fit tests adapted for right-censored data: \textit{Kolmogorov-Smirnov}, \textit{Anderson-Darling}, and \textit{Cramér-von Mises}. Through an extensive simulation study, we investigate how factors such as the proportion of right-censored observations, sample size, and distributional symmetry influence the statistical power of each test.

All procedures were implemented using the R package \texttt{GofCens}, and power estimates were obtained from extensive replicated simulations. The results identify the settings in which each test exhibits superior performance, providing practical, evidence-based recommendations for test selection in survival analysis. These recommendations are summarized as user-oriented guidelines and complemented by a concise graphical cheat sheet, enabling practitioners to quickly identify the most suitable test for their specific application.
\end{abstract}

\textbf{Keywords:} \texttt{GofCens} R package; Goodness-of-fit testing; Right-censored data; Survival analysis.

\section{Introduction}
\label{sec:intro}
Statistical models usually rely on certain assumptions about the underlying data-generating mechanism, among which the probability distribution of the observed data is one of the most fundamental. Assessing whether a theoretical distribution adequately represents the data is therefore a crucial step in statistical modeling. Accordingly, numerous goodness-of-fit tests have been developed for complete data and subsequently extended to accommodate right-censored observations.

Right-censored data arise naturally in survival analysis, biomedical research, and reliability engineering, where the event of interest is not observed for all subjects during the study period. Although nonparametric methods, such as the Kaplan-Meier estimator, are widely used because they require few distributional assumptions, parametric models remain essential for estimating hazard functions, extrapolating survival beyond the observed follow-up, and obtaining interpretable model parameters. Consequently, assessing the adequacy of a chosen parametric distribution is a fundamental step in time-to-event analysis.

Among the most widely used goodness-of-fit procedures are the \emph{Kolmogorov-Smirnov} (KS), \emph{Cramér-von Mises} (CvM), and \emph{Anderson-Darling} (AD) tests. These procedures evaluate the discrepancy between the empirical distribution function and the theoretical cumulative distribution function specified under the null hypothesis. 
Several extensions of these classical tests have been proposed for right-censored data. \citet{fleming1980modified} introduced a modified version of the KS test suitable for right-censored observations, while adaptations of the CvM and AD tests were developed by \citet{koziol76} and \citet{pettitt1976modified}, respectively.

Previous simulation studies comparing goodness-of-fit tests have focused predominantly on testing normality. For example, \citet{romao2010empirical} investigated the performance of several tests under different departures from normality, whereas \citet{pedrosa2015pruebas} evaluated their behaviour across a range of sample sizes. For right-censored data, \citet{kim2019tests} compared several procedures for assessing normality under different censoring mechanisms.

However, survival data rarely follow a normal distribution. Parametric survival models commonly assume distributions such as the Weibull, lognormal, gamma, log-logistic, or logistic, whose shapes can differ substantially in terms of skewness and tail behaviour. Despite the widespread use of these models, comparative evidence regarding the performance of goodness-of-fit tests under alternative distributions and varying censoring conditions remains limited. In particular, little attention has been paid to how distributional asymmetry influences the power of these tests.

To address this gap, this study presents a comprehensive Monte Carlo simulation comparing the KS, CvM, and AD goodness-of-fit tests for right-censored data under composite hypotheses with unknown parameters. The influence of sample size, censoring mechanism (administrative and random), censoring proportion, null distribution, and degree of skewness is systematically evaluated.  Furthermore, rather than presenting simulation results in isolation, we translate our empirical findings into practical, user-oriented decision rules. Specifically, we introduce the terminal survival discrepancy metric, $Diff_{t_{\max}}$, and integrate it into a structured selection protocol and flowchart designed to help practitioners select the most robust test for their specific data settings. Statistical power is estimated across a broad collection of simulation scenarios to provide practical guidance on the relative performance of each test.

All evaluated tests and diagnostic routines are fully implemented and accessible via the \texttt{R} package \texttt{GofCens} \citep{gofcensPackage, gofcensjournal}, which provides user-friendly graphical diagnostics and goodness-of-fit procedures for both complete and right-censored data across a wide range of probability distributions.

The remainder of this paper is organized as follows. Section~\ref{sec:Methods} presents the mathematical background, detailing the composite hypotheses, the right-censored test adaptations, the diagnostic terminal discrepancy metric, $\mathit{Diff}_{t_{\max}}$, and the quantile-based skewness measures used. Section~\ref{sec:Simulations} describes the design of the Monte Carlo simulation study and performance metrics. Section~\ref{sec:Results} reports the empirical results under no censoring, administrative censoring, and random censoring, as well as the behavior of the terminal survival discrepancy metric presented. Section~\ref{sec:Guidelines} synthesizes these findings into practical user guidelines, relying on the $\mathit{Diff}_{t_{\max}}$ diagnostic and the proposed decision flowchart. Finally, Section~\ref{sec:Discussion} offers concluding remarks, addresses study limitations, and outlines future research directions.
\section{Methodology}
\label{sec:Methods}
This section presents the methodology adopted in this study, focusing on the treatment of composite hypotheses, the formulation of the goodness-of-fit tests used, their adaptation to right-censored data, the introduction of  a diagnostic terminal survival discrepancy metric, $\mathit{Diff}_{t_{\max}}$, and the thresholds for how skewness is classified  for practical purposes.

\subsection{Notation and Hypothesis Formulation}
\label{subsec:Hypothesis}
Let $T_1, \dotsc, T_n$ be an independent and identically distributed random sample of latent failure times to an event of interest from an unknown cumulative distribution function (CDF) $F(t) = P(T \le t)$. In survival studies where right censoring occurs, let $C_1, \dotsc, C_n$ denote the corresponding censoring times, assumed to be independent of $T_i$. Rather than observing $T_i$ directly, the observed data for the $i$-th individual consist of the pair $(Y_i, \delta_i)$, defined as
\begin{equation*}
    Y_i = \min(T_i, C_i), \qquad 
    \delta_i = \mathds{1}(T_i \le C_i) = 
    \begin{cases} 
    1 & \text{if the event is observed,} \\ 
    0 & \text{if the observation is censored,} 
    \end{cases}
\end{equation*}
where $\mathds{1}(\cdot)$ represents the indicator function and $\delta_i$ is the event indicator.

The primary goal of goodness-of-fit testing is to assess whether the unknown underlying distribution $F$ belongs to a specified parametric family,
\begin{equation*}
  \mathcal{F}_0 = \{F_0(\cdot; \theta): \theta \in \Theta\},
\end{equation*}
where $F_0(\cdot;\theta)$ denotes the theoretical candidate CDF indexed by a parameter vector $\theta\in \Theta \subseteq \mathbb{R}^d$, and $\Theta$ is the  parameter space. Formally, we test the  composite null hypothesis, 
\begin{equation*}
    H_0: F(t) = F_0(t; \theta), \quad \text{for some } \theta \in \Theta,
\end{equation*}
against the general alternative $H_1: F(t) \notin \mathcal{F}_0$.

The goodness-of-fit tests evaluated in this work belong to the class of empirical distribution function (EDF) statistics, which quantify the discrepancy between an empirical nonparametric estimate of $F$ and the estimated theoretical distribution under $H_0$. For complete data, $F$ is consistently estimated by the classical sample EDF, $F_n(t) = \frac{1}{n} \sum_{i=1}^{n} \mathds{1}(T_i \leq t)$. In the presence of right censoring, unobserved event times prevent direct evaluation of $F_n(t)$. Instead, $F(t)$ is consistently estimated via its nonparametric maximum likelihood estimator, given by
\begin{equation*}
  \widehat{F}(t) = 1 - \widehat{S}(t) = 1 - \prod_{i: t_{(i)} \leq t} \Big( 1 - \frac{d_i}{n_i} \Big),
\end{equation*}
where $\widehat{S}(\cdot)$ denotes the Kaplan-Meier (KM) product-limit estimator \citep{Kaplan_Meier_1958}, $t_{(i)}$ represents the distinct ordered failure  times, $d_i$ is the number of events occurring at time $t_{(i)}$, and $n_i$ is the number of individuals at risk just prior to $t_{(i)}$. For ease of exposition, the explicit dependence on the sample size $n$ is omitted from the estimators $\widehat{F}(t)$, $\widehat{S}(t)$, $\hat{\theta}$, and $\widehat{\Lambda}(t)$ throughout the text.

Because the true parameter vector $\theta$ is unknown under $H_0$, it is estimated from the observed right-censored sample $(Y_i, \delta_i)_{i=1}^n$ using maximum likelihood estimation. The resulting estimate, denoted by $\hat\theta$, maximizes the right-censored likelihood function. Test statistics subsequently evaluate the distance between the estimated nonparametric distribution $\widehat{F}(t)$ and the fitted parametric model $F_0(t; \widehat{\theta})$.

\subsection{Formulation of Goodness-of-Fit Tests}
\label{subsec:Def_tests}
Three goodness-of-fit tests are considered in this study: the \textit{Kolmogorov-Smirnov}, \textit{Cramér-von Mises}, and \textit{Anderson-Darling} tests. Their formulations and adaptations to right-censored data are briefly reviewed below.

\subsubsection{Kolmogorov-Smirnov test} 
The \textit{Kolmogorov-Smirnov} (KS) statistic, is a supremum-based procedure that assesses goodness-of-fit by evaluating the maximum absolute vertical distance between the empirical and theoretical cumulative distribution functions. For complete, nonnegative failure time data ($T_i \ge 0$), the classical KS statistic under a composite null hypothesis is defined as
\begin{equation*}
    D_n = \sup_{t\geq 0} | F_n(t) - F_0(t; \widehat{\theta}) |,
\end{equation*}
where $F_n(t)$ is the standard sample EDF and $F_0(t; \widehat{\theta})$ is the fitted theoretical CDF with $\hat\theta$ the maximum likelihood estimator of $\theta$.

To accommodate right-censored observations, \cite{fleming1980modified} proposed an adapted KS statistic in terms of cumulative hazard functions. Let $\widehat{\Lambda}(t)$ denote the nonparametric Nelson-Aalen estimator of the cumulative hazard function \citep{Nelson_Aalen_1978}:
\begin{equation*}
  \widehat{\Lambda} (t)= \sum_{i: t_{(i)} \leq t}{\frac{d_i}{n_i}}, 
\end{equation*}
where, as earlier, $t_{(i)}$ represents the distinct ordered failure times, $d_i$ is the number of events occurring at $t_{(i)}$, and $n_i$ is the number of subjects at risk just prior to $t_{(i)}$. 
The right-censored version of the KS statistic is defined as
\begin{equation}
    \widehat{D}_n = \sup_{0 \leq t \leq t_{max}} \big| \int_{0}^{t} \frac{\widehat{S}(t)S_0(s;\hat\theta)}{\widehat{S}(s)} d[\widehat{\Lambda}(s) - \Lambda_0(s;\hat\theta)] \big|
    \label{eq:KS_modif}
\end{equation}
where $t_{max}= \max_{1 \le i \le n} Y_i$ represents the maximum  observed time in the sample, $\widehat{S}(\cdot)$ is the Kaplan-Meier estimator of the survival function, and $S_0(\cdot; \widehat{\theta})$ and $\Lambda_0(\cdot; \widehat{\theta})$ denote the theoretical survival and cumulative hazard functions under $H_0$, respectively.

Computing $D_n$ directly via the raw vertical  differences $|\widehat{F}(t) - F_0(t; \widehat{\theta})|$ would introduce extreme volatility at the upper tail of the distribution, where very few individuals remain at risk and the KM estimator loses precision. The adaptation of \cite{fleming1980modified} in \eqref{eq:KS_modif}, resolves this instability by shifting the comparison from the direct CDFs to the cumulative hazard functions. By introducing the weighting fraction $\frac{\widehat{S}(t)S_0(s;\hat\theta)}{\widehat{S}(s)}$ into the integrand, the statistic dynamically stabilizes the variance. This ensures that a random jump at the very end of a survival curve (where sample size effectively drops) does not artificially inflate the maximum distance and distort test conclusions.

\subsubsection{Cramér-von Mises \& Anderson-Darling tests}
Unlike the supremum-based \textit{Kolmogorov-Smirnov} test, the \textit{Cramér-von Mises} (CvM) and \textit{Anderson-Darling} (AD) tests belong to the family of quadratic empirical distribution function statistics. Rather than focusing on the maximum absolute deviation between the empirical and theoretical distribution functions, these procedures measure the integrated squared discrepancy between the empirical and theoretical cumulative distribution functions over the entire support of the distribution. The classical CvM statistic for complete data is defined as
\begin{equation*}
    M_n = n \int_{-\infty}^{+\infty} \big(F_n(t) - F_0(t;\hat\theta)\big)^2 dF_0(t;\hat\theta).
\end{equation*}

The AD test can be viewed as a weighted version of the CvM statistic, where the integrand is weighted inversely proportional to the variance of the empirical distribution: 
\begin{equation*}
  A_n = n \int_{-\infty}^{+\infty} \frac{\big(F_n(t) - F_0(t;\hat\theta)\big)^2}{F_0(t;\hat\theta)\big(1 - F_0(t;\hat\theta)\big)} dF_0(t;\hat\theta).
\end{equation*}
This weighting places  greater emphasis on discrepancies occurring in the lower and upper tails of the  candidate distribution.

For right-censored data, we adopt the extension proposed by \citet{pettitt1976modified} and \citet{koziol76}. Let $Y_{(1)}, \dotsc, Y_{(n_r)}$ represent the ordered observed failure times, where $n_r = \sum_{i = 1}^{n}{\delta_i}$ denotes the total number of uncensored events. These failure times are transformed to the uniform scale via $u_{(i)} = F_0(Y_{(i)};\hat\theta)$ for $j=1, \dotsc, n_r$, subject to the boundary conventions $u_{(0)} = 0$ and $u_{(n_r + 1)} = 1$. The adapted \textit{Cramér-von Mises} statistic is defined algebraically as:
\begin{equation*}
  \widehat{M}_n = n_r \sum_{j=1}^{n_r+1} F_n(u_{(j-1)}) (u_{(j)} - u_{(j-1)}) \big(F_n(u_{(j-1)}) - (u_{(j)} + u_{(j-1)}) \big) + \frac{n_r}{3}.
\end{equation*}

Similarly, the adapted \textit{Anderson-Darling} statistic $\widehat{A}_n$ is computed using the weighted differences between the estimated survival curve and the null model at each observed failure time as:
\begin{equation}
    \begin{split}
        \widehat{A}_n = -n_r &+ n_r\sum_{j = 1}^{n_r}\big(F_n(u_{(j - 1)}) - 1\big)^2 \big(\log(1 - u_{(j-1)}) - \log(1 - u_{(j)})\big)\\
        &+ n_r\sum_{j=1}^{n_r-1}F_n^2(u_{(j)})\big(\log (u_{(j+1)}) - \log(u_{(j)})\big)-n_r\log(u_{(n_r)}).
    \end{split}
    \label{eq:AD_modif}
\end{equation}

The main difference between the classical EDF statistics and their right-censored counterparts lies in how discrepancies are accumulated. For complete data, the statistics integrate the distance between the empirical and theoretical distribution functions over the entire support. Under right censoring,  upper-tail information  is only partially observed, making the classical integral representation unfeasible. To overcome this limitation, the adapted statistics introduce two fundamental modifications. First, the continuous integral over $\mathbb{R}$ is replaced by a discrete sum that evaluates the discrepancy exclusively at observed failure times,  weighting each term by ($u_{(j)} - u_{(j-1)}$). Second, the scaling normalization factor changes from the total sample size $n$ to the number of observed events $n_r$.

Although the \textit{Anderson-Darling} statistic is designed to emphasize discrepancies in the tails of the distribution, right censoring substantially reduces the information available in this region. The modified discrete formulation in \eqref{eq:AD_modif} preserves the weighting structure of the original statistic while restricting the calculation to intervals determined by the observed failure times, thereby reducing the instability that would arise from evaluating heavily censored portions of the distribution. However, as shown in Section~\ref{sec:Results}, severe administrative censoring near $u_{(n_r)}$ can reduce the reliability of these statistics by limiting the information available in the upper tail, highlighting the diagnostic importance of terminal discrepancies.

All three goodness-of-fit procedures (KS, CvM, and AD) under right censoring are fully implemented in the \texttt{R} package  \texttt{GofCens} via the functions \texttt{KScens()}, \texttt{CvMcens()}, and \texttt{ADcens()} \citep{gofcensPackage, gofcensjournal}. Because censoring and composite parameter estimation distort standard limiting distributions, p-values are obtained using parametric bootstrap resampling.

\subsection{Terminal Survival Discrepancy Metric: $\mathit{Diff}_{t_{\max}}$}
\label{subsec:Diff}
While classical EDF test statistics aggregate discrepancies across the observed follow-up period via integration or supremum operations, they can obscure localized structural misfits occurring in the extreme right tail. That is, systematic departures between the assumed parametric family and the true distribution that are confined to a specific region of the support, particularly the upper tail, may go undetected. Integrated tests such as CvM sum discrepancies across the entire follow-up period and may attenuate the contribution of localized tail divergences. Because relatively few individuals remain at risk in the extreme tail, and censoring may be substantial, a substantial discrepancy near ($t_{\max}$) may be outweighed by a good fit over much of the earlier follow-up period, potentially resulting in a large, noninformative p-value. Tail-weighted tests such as AD are more sensitive to tail discrepancies, although their ability to detect an extreme-tail departure may still be limited by the amount of information available in that region. On the other hand, for supremum-based tests such as KS, the  maximum distance $D_n$ may instead be attained elsewhere, so a localized structural divergence developing late in follow-up need not determine the test statistic.

This limitation highlights the value of a targeted tail diagnostic as a complement to global goodness-of-fit test statistics. Furthermore, under right censoring, upper-tail behavior is particularly vulnerable to estimation uncertainty caused by the limited number of observations and substantial loss to follow-up. To capture this specific aspect, we introduce a diagnostic metric, denoted by $\mathit{Diff}_{t_{\max}}$, which measures the signed vertical discrepancy between the fitted null model and the nonparametric estimator at the terminal observed follow-up time. This metric is not intended as an additional test statistic, but rather as a complementary diagnostic that, as explained in the following sections, can help assess which goodness-of-fit test may be most informative for a given data set.

\begin{figure}
\centering   
\includegraphics[width=0.65\linewidth]{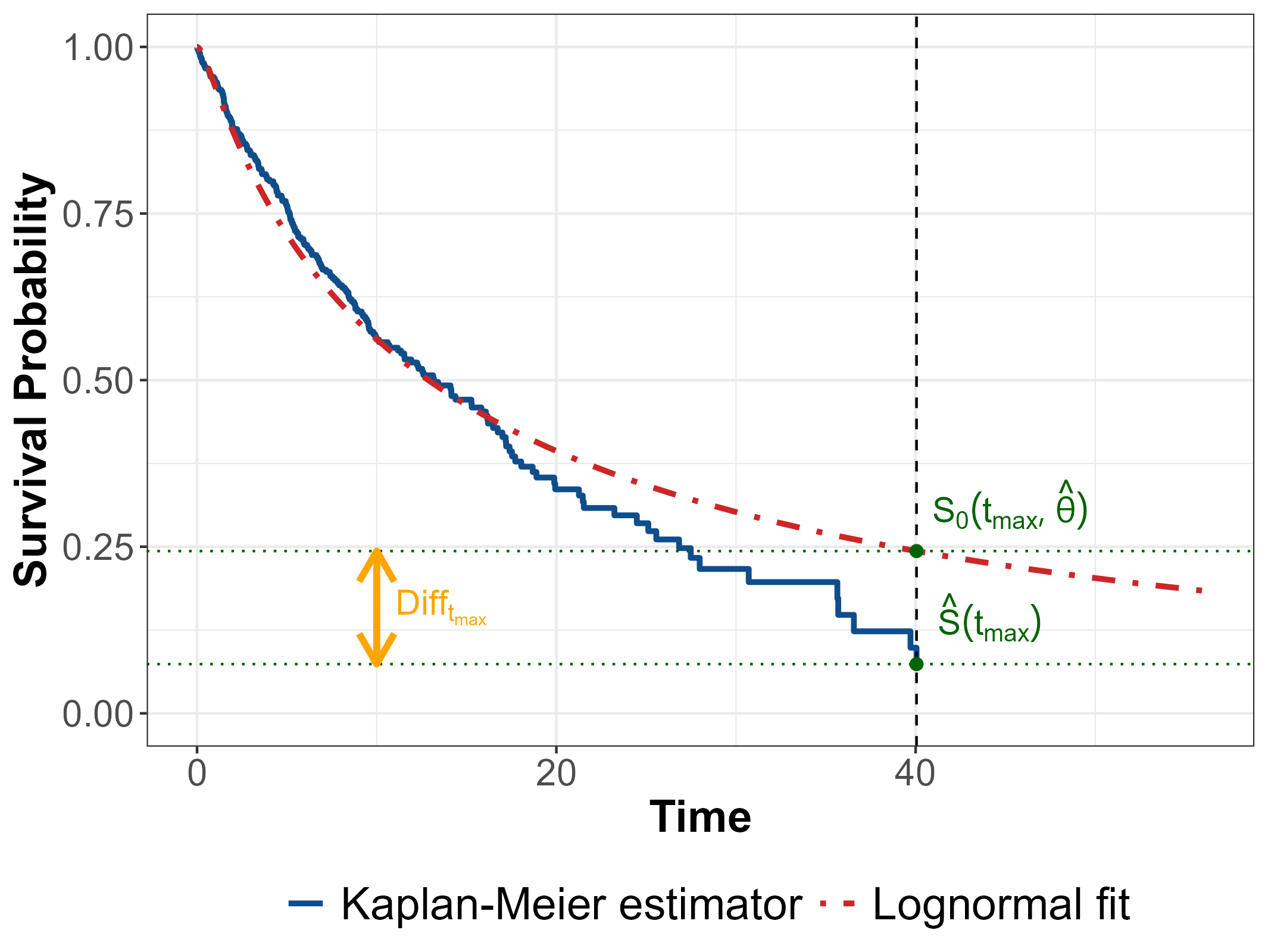}   
\caption{Schematic representation of the terminal survival discrepancy metric $\mathit{Diff}_{t_{\max}}$. It is defined as the difference between the parametric fit and the Kaplan-Meier estimator of the survival function at the maximum observed follow-up time $t_{\max}$; see equation~\eqref{eq:diffmetric}.}
\label{fig:Figure1}
\end{figure}

Let $t_{\max} = \max_{1 \le i \le n} Y_i$ represent the maximum observed time in the sample. The terminal discrepancy metric is defined as
\begin{equation}
\label{eq:diffmetric}
  \mathit{Diff}_{t_{\max}} = S_0(t_{\max}; \widehat{\theta}) - \widehat{S}(t_{\max}),
\end{equation}
where $\widehat{S}(t_{\max})$ is the Kaplan–Meier estimator of the survival function evaluated at $t_{\max}$, and $S_0(t_{\max}; \widehat{\theta})$ is the theoretical survival probability under the null hypothesis, evaluated at the maximum likelihood estimate $\widehat{\theta}$. A graphical representation of this measure is shown in Figure~\ref{fig:Figure1}.

By design, $\mathit{Diff}_{t_{\max}}$ retains its algebraic sign rather than using its absolute magnitude ($|\mathit{Diff}_{t_{\max}}|$), because the sign provides important directional diagnostic information:
\begin{itemize}
    \item A \textbf{positive discrepancy} ($\mathit{Diff}_{t_{\max}} > 0$) indicates that the fitted parametric survival function lies above the Kaplan-Meier estimator at $t_{\max}$.
    \item A \textbf{negative discrepancy} ($\mathit{Diff}_{t_{\max}} < 0$) indicates that the fitted parametric survival function lies below the Kaplan-Meier estimator at $t_{\max}$.
\end{itemize}

\subsection{Quantile-Based Skewness Measures}
\label{subsec:Skewness}
Quantifying the degree of asymmetry in the candidate distributions is essential for constructing controlled simulation scenarios across diverse shape parameters. This allows a systematic analysis of how distributional skewness  impacts the statistical power and behavior of goodness-of-fit tests under right censoring.

In survival and reliability settings, classical moment-based skewness coefficients (such as the standardized third central moment) are often unreliable or undefined because right censoring prevents complete observation of the tail. Moreover, defining skewness based solely on the parametric null distribution would introduce model dependence into the analysis, which is undesirable in the context of goodness-of-fit assessment, where the underlying distribution is precisely the object being evaluated. Consequently, this study considers robust, nonparametric, quantile-based skewness measures.

The primary measure considered is Bowley’s coefficient of skewness \citep{Bowley1901}, also known as Yule’s coefficient \citep{Yule1912}, which quantifies asymmetry based on the interquartile range:
\begin{equation*}
  A_Y = \frac{(Q_3-Q_2)-(Q_2-Q_1)}{Q_3-Q_1},
\end{equation*}
where $Q_1$, $Q_2$ (the median), and $Q_3$ denote the $25^\mathrm{th}$, $50^\mathrm{th}$, and $75^\mathrm{th}$ percentiles, respectively.

To capture asymmetry over a wider portion of the distribution, Kelly's coefficient of skewness is also considered:
\begin{equation*}
  A_K = \frac{(P_{90}-Q_2)-(Q_2-P_{10})}{P_{90}-P_{10}},
\end{equation*}
where $P_{10}$ and $P_{90}$ denote the $10^\mathrm{th}$ and $90^\mathrm{th}$ percentiles, respectively. Under right censoring, the percentiles $P_p$ are estimated nonparametrically from the Kaplan--Meier survival curve by solving
\begin{equation*}
\widehat{S}(P_p) = 1-\frac{p}{100},
\end{equation*}
when the corresponding percentile is estimable, where $p$ denotes the percentile as a percentage.

Both coefficients take values in the interval $[-1,1]$. A value of zero indicates symmetry, while positive and negative values indicate right- and left-skewed distributions, respectively. Throughout this manuscript, for practical purposes, the degree of distributional asymmetry is categorized according to the absolute value of the skewness coefficient, $|A|$:
\begin{enumerate}
  \item Symmetric: $|A| \in [0,0.1)$.
  \item Moderately asymmetric: $|A| \in [0.1,0.4)$.
  \item Extremely asymmetric: $|A| \in [0.4,1]$.
\end{enumerate}

Evaluating Bowley's and Kelly's coefficients requires observing events up to $Q_3$ ($75^\mathrm{th}$ percentile) and $P_{90}$ ($90^\mathrm{th}$ percentile), respectively. Under moderate or heavy censoring, these higher-order quantiles may not be estimable if the largest observed event time occurs before the required percentile. To accommodate such scenarios, we consider a generalized quantile-based skewness coefficient:
\begin{equation*}
  A_g(p) = \frac{(P_{100-p}-Q_2)-(Q_2-P_p)}{P_{100-p}-P_p},
\end{equation*}
where $P_p$ and $P_{100-p}$ represent the lower and upper quantiles of orders $p$ and $100-p$, respectively, with $p\in(0,50)$ chosen according to the maximum observable follow-up time $t_{\max}$.

While this generalized coefficient $A_g(p)$ extends skewness evaluation to settings with heavier censoring, it should be applied with caution: it should only be used when both selected quantiles can be estimated with sufficient reliability. Under severe tail censoring, the available quantile range may become too narrow to adequately reflect the true asymmetry of the distribution, potentially leading to an underestimation of skewness.

\section{Simulation Study Design}
\label{sec:Simulations}
The purpose of this simulation study is to evaluate the performance of the previously presented goodness-of-fit tests adapted to right-censored data in survival analysis. We investigate how key factors, such as sample size, the underlying distribution, skewness, and the proportion of right-censored observations, affect the power of these tests. The results, presented in Section~\ref{sec:Results}, are summarized using relevant performance metrics to provide practical guidance for selecting the most appropriate goodness-of-fit test for a given data setting.

\subsection{Simulation Settings}
\label{subsec:Settings}
\begin{figure}
    \centering
    \includegraphics[width=1\textwidth]{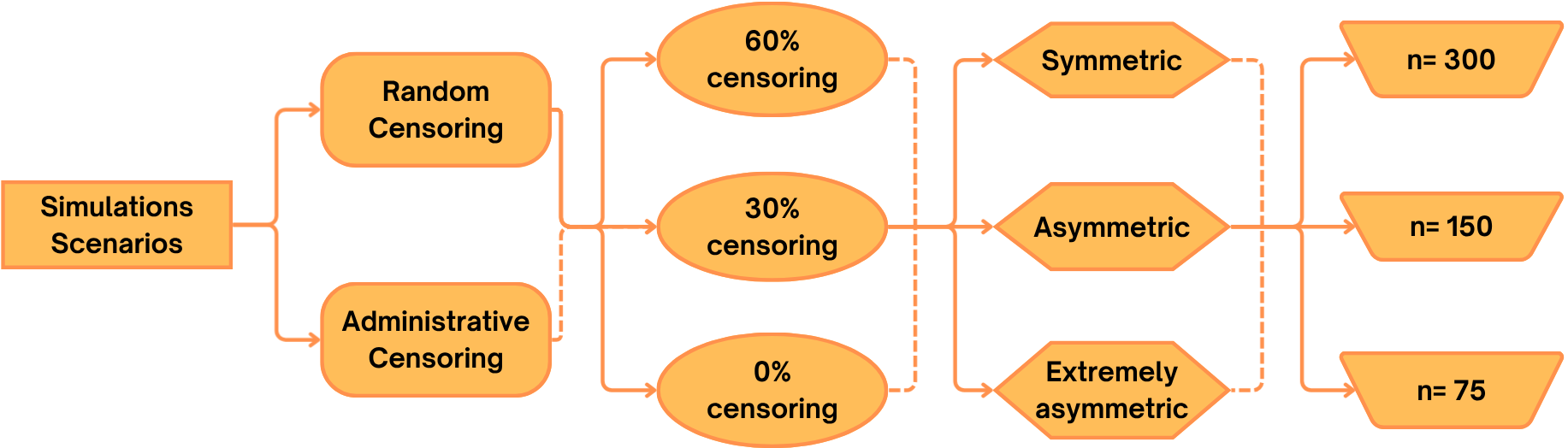}
    \caption{Summary diagram of the simulation settings considered for the Weibull and the lognormal distributions.}
    \label{fig:Figure2}
\end{figure}

The simulation study considers several factors that may affect the performance of goodness-of-fit tests for right-censored data. Specifically, we vary the data-generating distribution, degree of skewness, censoring mechanism, proportion of censored observations, and sample size. The combinations of these factors define the simulation settings for both data-generating distributions, which are summarized in Figure~\ref{fig:Figure2}.

\paragraph{Data-generating distributions.}
Data are generated from two distributions commonly used in survival analysis: the Weibull distribution with shape and scale parameters $\alpha$ and $\beta$, and the lognormal distribution with location parameter $\mu$ and scale parameter $\beta$. These distributions allow for different distributional shapes and hazard behaviors. In particular, the Weibull distribution can accommodate increasing, decreasing, or constant hazard functions depending on its shape parameter, whereas the lognormal distribution can represent right-skewed time-to-event data. Their corresponding survival functions are given by
\begin{align*}
    S_{\text{W}}(t)  &= e^{-(\frac{t}{\beta})^\alpha}, \\[1ex]
    S_{\text{LN}}(t) &= \int_{\frac{\log t - \mu}{\beta}}^\infty \frac{1}{\sqrt{2\pi}} e^{-\frac{x^2}{2}} \, dx.
\end{align*}

\paragraph{Degree of skewness.}
For each data-generating distribution, three levels of skewness are considered: symmetric, moderately skewed, and extremely skewed. These settings are obtained by specifying appropriate parameter values for the data-generating distributions, as reported in Table~\ref{tab:Data_gen_param}. The parameter values were initially selected based on graphical inspection and subsequently validated using the Yule and Kelly skewness coefficients described in Section~\ref{subsec:Skewness}. The corresponding coefficients computed from the simulated data are also reported in Table~\ref{tab:Data_gen_param}.

\renewcommand{\arraystretch}{1.5}
\begin{table}[H]
  \centering
  \begin{tabular}{lcccccccc}
    \cline{2-9}
      & \multicolumn{4}{c}{Weibull} & \multicolumn{4}{c}{Lognormal} \\
    \cmidrule(rl){2-5} \cmidrule(lr){6-9}
      & $\alpha$ & $\beta$ & Yule's & Kelly's & $\mu$ & $\beta$ & Yule's & Kelly's \\
    \midrule
      Symmetric         & 3.5 & 20 & 0.007 & 0.008 & 2 & 0.01 & 0.003 & 0.006 \\
      Moderately Skewed & 1.7 & 20 & 0.109 & 0.210 & 2 & 0.4  & 0.134 & 0.251 \\
      Extremely Skewed  & 1   & 20 & 0.262 & 0.465 & 2 & 1.5  & 0.467 & 0.745 \\
    \bottomrule
  \end{tabular}
\caption{Parameter settings for the Weibull$(\alpha,\beta)$ and Lognormal$(\mu,\beta)$ distributions under each skewness setting, together with the corresponding Yule and Kelly skewness coefficients computed from the simulated data.}
\label{tab:Data_gen_param}
\end{table}

\paragraph{Censoring mechanism.}
Two censoring mechanisms are considered: random censoring and administrative censoring. Random censoring represents a common setting in survival analysis in which censoring may arise from loss to follow-up or from individuals entering the study at different times. Administrative censoring, in contrast, occurs when all individuals are subject to the same fixed censoring time corresponding to the end of follow-up.

For random censoring, censoring times are generated independently from an exponential distribution with scale parameter $\beta$. This parameter is adjusted according to the data-generating distribution, degree of skewness, and target censoring proportion. The values selected to achieve the desired censoring proportions are reported in Table~\ref{tab:Exponential_param}. Administrative censoring does not require an additional censoring distribution, as all individuals are censored at the same fixed time point.

\paragraph{Censoring proportion.}
Three levels of right censoring are considered: $0\%$, $30\%$, and $60\%$. The $0\%$ setting corresponds to complete data and provides a benchmark for assessing the effect of censoring. The $30\%$ and $60\%$ settings correspond, respectively, to approximately $70\%$ and $40\%$ of observations remaining uncensored.

\begin{table}[H]
  \centering
  \begin{tabular}{lcccc}
    \cline{2-5}
      & \multicolumn{2}{c}{30\% Censoring} & \multicolumn{2}{c}{60\% Censoring} \\
    \cmidrule(rl){2-3} \cmidrule(lr){4-5}
      & Weibull & Lognormal & Weibull & Lognormal \\
    \midrule
      Symmetric         & $48.9$  & $20.5$  & $18.5$  & $7.99$ \\
      Moderately Skewed & $46.28$ & $21.55$ & $15.95$ & $8.01$ \\
      Extremely Skewed  & $45.9$  & $32.5$  & $13.2$  & $7.25$ \\
    \bottomrule
  \end{tabular}
\caption{Scale parameters $\beta$ of the exponential distributions used to generate random censoring for each combination of censoring proportion, data-generating distribution, and degree of skewness.}
\label{tab:Exponential_param}
\end{table}

The exponential scale parameters reported in Table~\ref{tab:Exponential_param} were determined separately for each combination of data-generating distribution and skewness level by simulation, selecting the values that most closely matched the target censoring proportions.

\paragraph{Sample size.}
Three sample sizes are considered: $n=75$, $n=150$, and $n=300$. These values allow the performance of the tests to be assessed across small, moderate, and relatively large samples.

\paragraph{Overall simulation design.}
In total, the number of simulation settings, defined by the combinations of data-generating distribution, skewness degree, censoring proportion, censoring mechanism, and sample size, is
$2 \times 3 \times 3 \times 2 \times 3 = 108$. For each setting, $N=1000$ samples are generated. Each goodness-of-fit test is then applied to every simulated sample under each of the three hypothesized null distributions, and empirical power is computed for each combination of simulation setting and null hypothesis.

The three hypothesized null distributions are the Weibull, lognormal, and logistic distributions. Thus, although event times are generated from either a Weibull or a lognormal distribution, all three candidate distributions are considered as null hypotheses for each simulated dataset. This design allows us to assess test performance when the hypothesized distribution is either the true data-generating distribution or a competing alternative.

\subsection{Performance Metrics}
\label{subsec:Metrics}
To benchmark and compare the empirical performance of the goodness-of-fit tests across the simulation scenarios, we evaluate two fundamental criteria: empirical Type I error rate control under the null hypothesis and statistical power under alternative distributions.

Recall that $F$ denotes the true data-generating cumulative distribution function (CDF), and $\mathcal{F}_0 = \{F_0(\cdot; \theta) : \theta \in \Theta\}$ represents the candidate parametric family under evaluation, where $\Theta \subseteq \mathbb{R}^d$ is the parameter space. The composite goodness-of-fit hypothesis testing problem is stated as
\begin{equation}
    H_0: F \in \mathcal{F}_0 \quad \text{versus} \quad H_1: F \notin \mathcal{F}_0,
    \label{eq:hypothesis_testing}
\end{equation}
or, equivalently, in terms of the candidate distribution function:
\begin{equation*}
    H_0: F(t) = F_0(t; \theta) \text{ for some } \theta \in \Theta \quad \text{versus} \quad H_1: F(t) \neq F_0(t; \theta) \text{ for all } \theta \in \Theta.
\end{equation*}

The power function of a statistical test \citep[][Section~3.1]{lehmann2005}, denoted by $\pi(F)$, defines the probability of rejecting $H_0$ given that the data are generated by $F$:
\begin{equation}
    \pi(F) = \mathbb{P}_{F}\left( \text{Reject } H_0 \right).
    \label{eq:power_function}
\end{equation}
The interpretation of $\pi(F)$ depends directly on whether $F$ satisfies the null condition:
\begin{itemize}
  \item \textbf{Type I Error Rate ($F \in \mathcal{F}_0$):} When $F$ belongs to the null parametric family $\mathcal{F}_0$, $\pi(F)$ represents the probability of a false positive result. For a test specified at nominal significance level $\alpha \in (0,1)$, size validity requires that $\sup_{F \in \mathcal{F}_0} \pi(F) \le \alpha$.

  \item \textbf{Statistical Power ($F \notin \mathcal{F}_0$):} When $F$ lies outside the null family $\mathcal{F}_0$, $\pi(F)$ measures the power of the test, representing the probability of correctly rejecting an inadequate parametric model.
\end{itemize}

In our Monte Carlo study, the true rejection probability $\pi(F)$ is estimated empirically by generating $N = 1000$ independent right-censored samples under each specified simulation condition:
\begin{equation*}
  \label{eq:empirical_power}
  \widehat{\pi}(F) = \frac{1}{N} \sum_{i=1}^{N} \mathds{1}(p_i \le \alpha),
\end{equation*}
where $p_i$ is the parametric bootstrap $p$-value obtained from the $i^\mathrm{th}$ simulated sample, $\mathds{1}(\cdot)$ is the indicator function, and $\alpha = 0.1$ is the nominal significance level fixed throughout this work.

Under $H_0$ ($F \in \mathcal{F}_0$), the Monte Carlo standard error of the empirical Type I error estimator is given by $\text{SE}(\widehat{\pi}) = \sqrt{\alpha(1 - \alpha) / N} \approx 0.0095$ for $\alpha = 0.1$ and $N = 1000$. Consequently, an empirical Type I error rate falling within the $95\%$ binomial confidence interval $[0.081, 0.119]$ indicates adequate nominal level control, whereas values exceeding $0.119$ signify liberal behavior (inflated Type I error) and values below $0.081$ indicate conservative test behavior. Under alternative settings ($F \notin \mathcal{F}_0$), $\widehat{\pi}(F)$ provides an unbiased estimate of statistical power, where higher values reflect greater sensitivity to structural distributional misspecification.

Although empirical power allows us to identify the best-performing test under each simulation setting, such a choice is not straightforward in real-data applications, where the underlying distribution and other characteristics of the data are typically unknown. In practice, only the observed, possibly censored, data are available, and therefore the selection of an appropriate goodness-of-fit test must rely on information that can be obtained directly from the observed sample. For this reason, the terminal survival discrepancy metric, $Diff_{t_{\max}}$, is recorded and investigated as a potential decision-making tool for test selection. This measure provides a simple and readily estimable summary of the discrepancy between the nonparametric and fitted parametric survival functions at $t_{\max}$, and may provide information about characteristics of the data that are relevant to the performance of the goodness-of-fit tests.

To account for the variability across samples within each simulation setting, $Diff_{t_{\max}}$ is computed for each simulated sample as the difference between $\widehat{S}_{H_0}(t_{\max}, \widehat{\theta})$ and $\widehat{S}(t_{\max})$. The resulting values are then summarized across the $N=1000$ simulated samples for each setting. Together with the corresponding $p$-values, these quantities are used to characterize the performance of the tests across the different simulation settings and to investigate whether $Diff_{t_{\max}}$ can provide practical guidance for test selection when the underlying distribution is unknown. The results of these analyses are presented in Section~\ref{sec:Results}.

\section{Simulation Study Results}
\label{sec:Results}
The results of the simulation study will be presented graphically as follows. For each test, its performance will be evaluated according to the underlying data-generating distribution, considering each null hypothesis and all simulated scenarios. The results will be presented separately for uncensored data, administrative censoring, and random censoring. Figures not included in this section, are shown in the Appendix~\ref{sec:appA}.

\subsection{Uncensored Data}
\label{subsec:Complete_data}
\begin{figure}
\centering
  \includegraphics[width=1\textwidth]{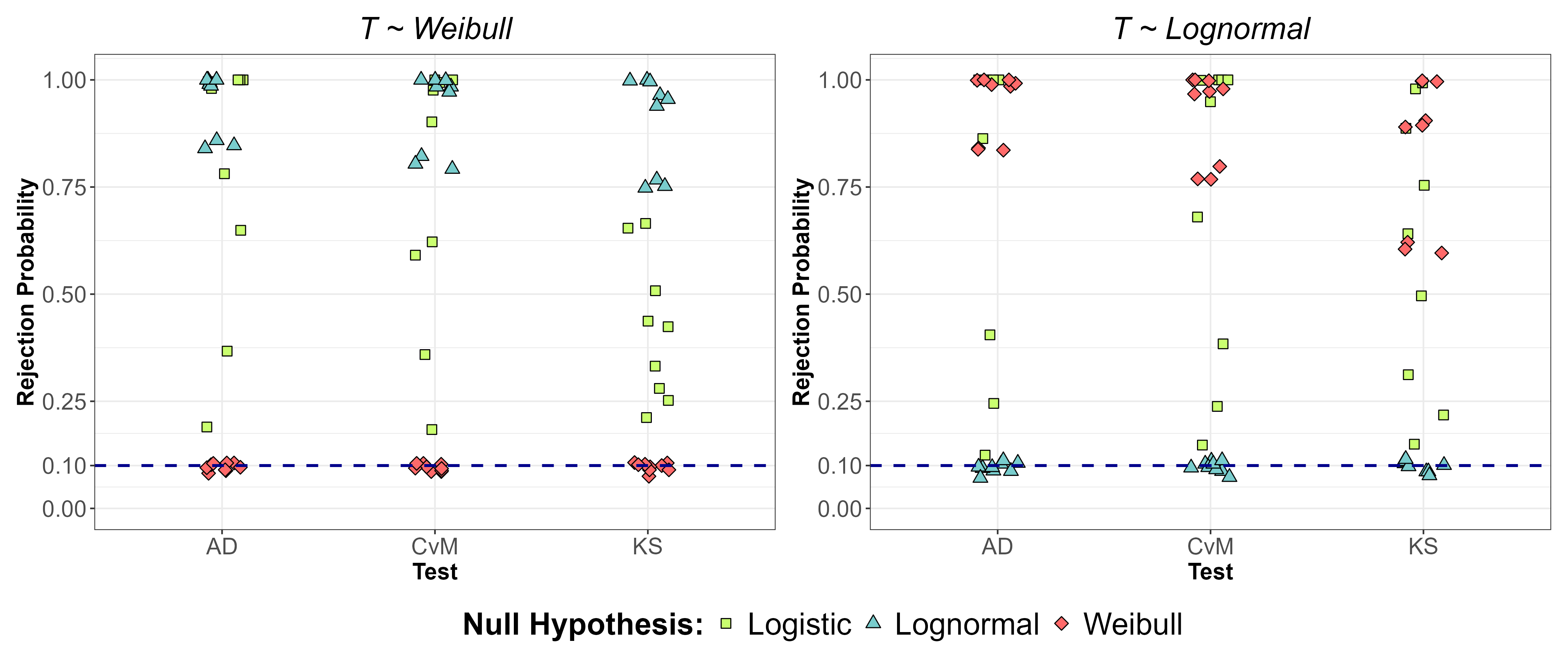}
  \caption{Comparative plots of the estimated rejection probabilities of goodness-of-fit tests (\textit{Anderson-Darling} (AD), \textit{Cramér-von Mises (CvM)}, and \textit{Kolmogorov-Smirnov} (KS), in that order) for uncensored data under different null-hypothesis distributions: logistic (green squares), lognormal (blue triangles), and Weibull (red diamonds). The left panel shows the behavior of the tests for data generated from a Weibull distribution, while the right panel presents the results for data generated from a lognormal distribution. In both panels, for each test and null hypothesis, each point corresponds to a different combination of sample size and degree of skewness. Horizontal jitter has been added to the points to improve the visual separation between overlapping observations.}
  \label{fig:fig3}
\end{figure}

Figure \ref{fig:fig3} illustrates the results for the scenarios with complete observations. Across all tests and scenarios, the Type I error appears to be well controlled at the nominal $0.1$ level. In both panels, when the data-generating distribution coincides with the null hypothesis, the empirical rejection rates are close to $0.1$, as expected for tests conducted at a nominal significance level of $\alpha = 0.1$. This indicates that the tests are reasonably well calibrated under the null hypothesis. Proper control of the Type I error is particularly important, as it provides a basis for interpreting differences in power across the remaining scenarios.

For data generated from either of the two distributions, the tests generally have difficulty distinguishing the data from a logistic distribution. In contrast, all tests show higher power to distinguish between the Weibull and lognormal distributions. When comparing the individual tests, the \textit{Anderson-Darling} test consistently exhibits the highest power in nearly all scenarios. In contrast, the \textit{Kolmogorov-Smirnov} test systematically shows the lowest power, while the \textit{Cramér-von Mises} test performs reasonably well across the different scenarios. 

Although the effect of sample size is not shown explicitly in the figure, the results also indicate the expected increase in power as the sample size increases. The role of the degree of skewness will be discussed in the Section~\ref{subsec:Rand_cens}.

\subsection{Administrative Censored Data}
\label{subsec:Admin_cens}
\begin{figure}
\centering
  \includegraphics[width=0.8\linewidth]{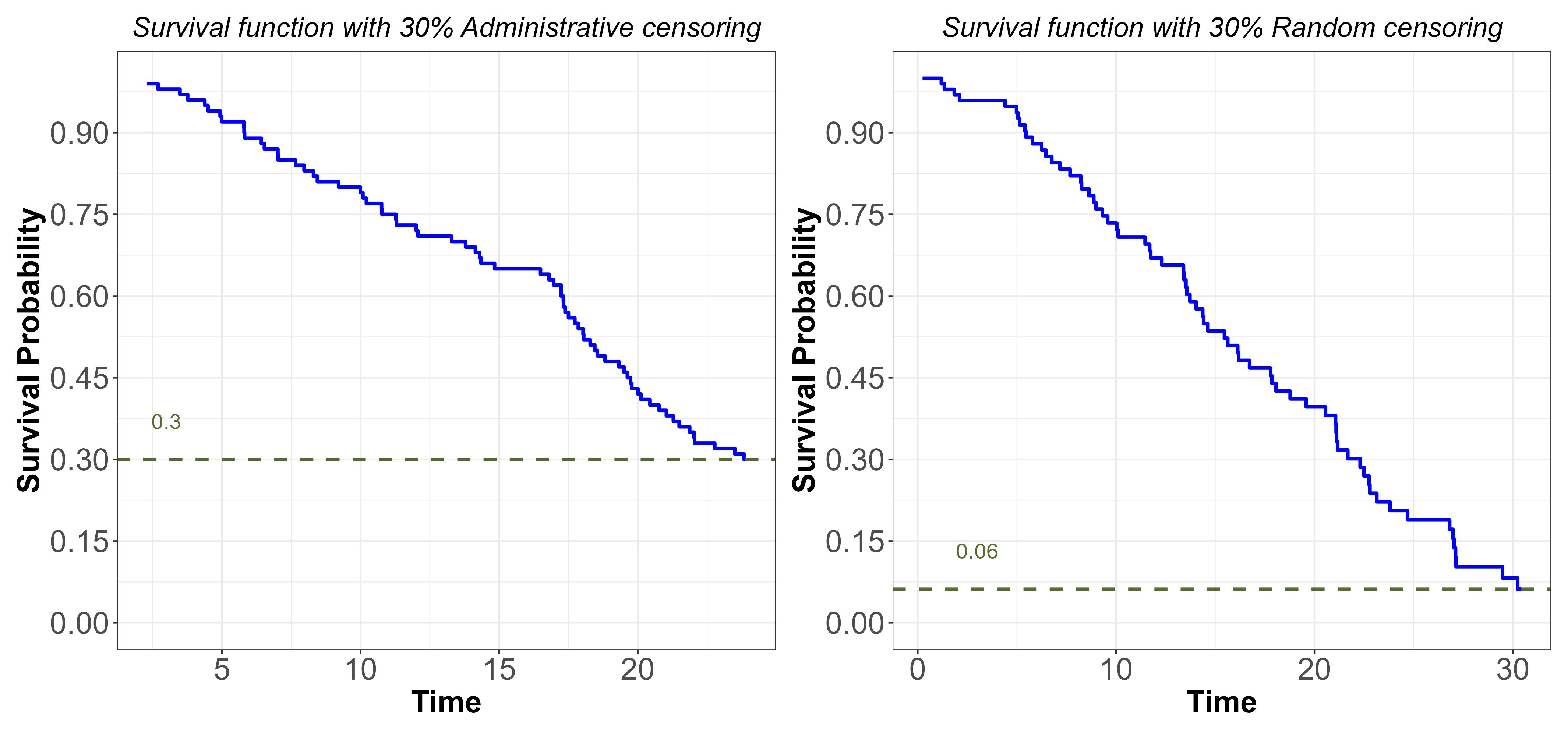}
  \caption{Illustration of Kaplan-Meier curves with 30\% of administrative censoring (left panel) and 30\% of random censoring (right panel).}
  \label{fig:fig4}
\end{figure}

An example of administrative censoring at a fixed censoring proportion of $30\%$ data is shown in the left panel of  Figure~\ref{fig:fig4}, which displays the Kaplan-Meier estimate of the survival curve for one such dataset.

\begin{figure}
\centering
  \includegraphics[width=1\linewidth]{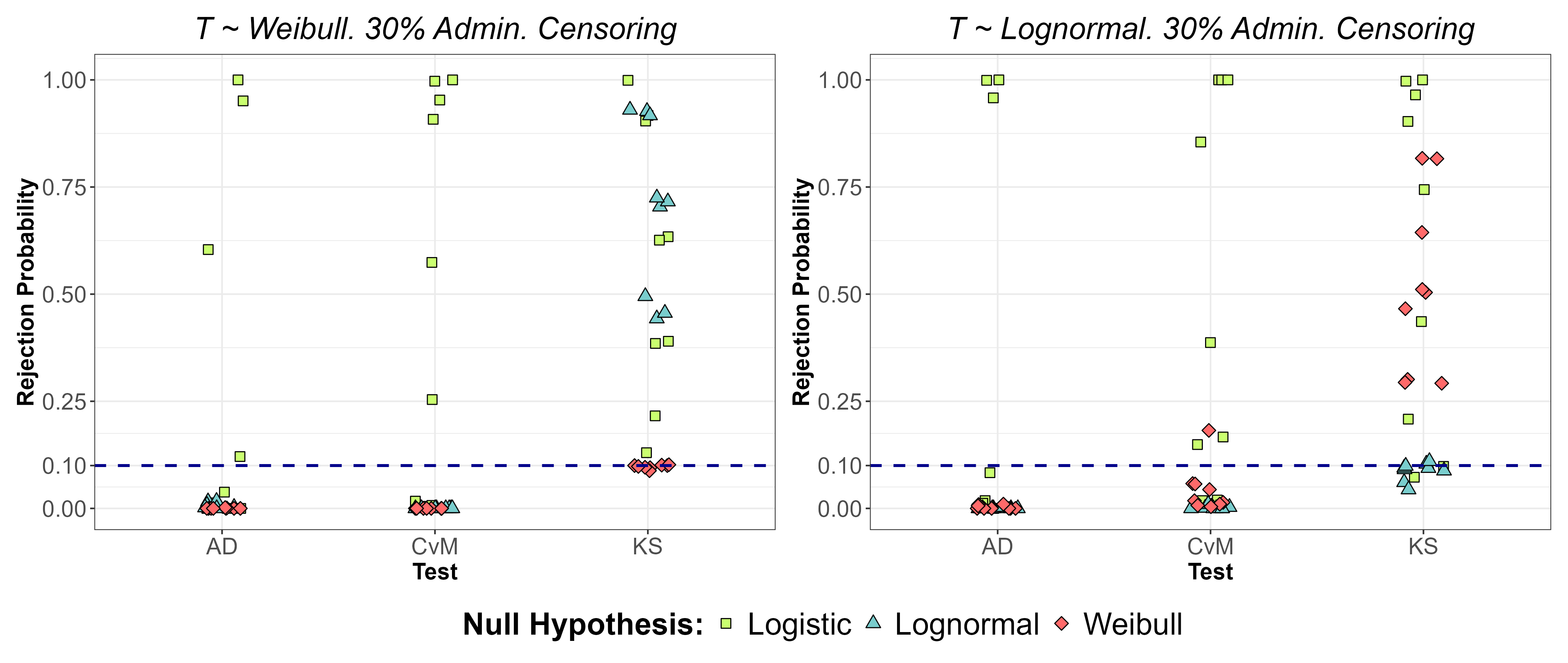}
  \caption{Comparative plots of the estimated rejection probabilities of goodness-of-fit tests (\textit{Anderson-Darling} (AD), \textit{Cramér-von Mises (CvM)}, and \textit{Kolmogorov-Smirnov} (KS), in that order) for 30\% administrative censored data under different null-hypothesis distributions: logistic (green squares), lognormal (blue triangles), and Weibull (red diamonds). The left panel shows the behavior of the tests for data generated from a Weibull distribution, while the right panel presents the results for data generated from a lognormal distribution. In both panels, for each test and null hypothesis, each point corresponds to a different combination of sample size and degree of skewness. Horizontal jitter has been added to the points to improve the visual separation between overlapping observations.}
  \label{fig:fig5}
\end{figure}

In scenarios where censoring truncates the tails of the data abruptly, the tests tend to exhibit substantially reduced rejection probabilities. Focusing on the \textit{Anderson-Darling} and \textit{Cramér-von Mises} tests in Figures~\ref{fig:fig5} (30\% administrative censoring) and~\ref{fig:figA1} (60\% administrative censoring), we observe that their empirical Type I error rates are generally below the nominal significance level of $\alpha=0.1$, indicating conservative behavior. This conservative behavior is accompanied by reduced power, making these tests less able to distinguish between the data-generating distributions and the null hypotheses. In particular, even when the null hypothesis coincides with the true data-generating distribution, the empirical rejection probability remains below the nominal level.

This behavior can be understood by noting that both tests rely on the discrepancy between the estimated survival curves. When the tails are heavily truncated, the loss of tail information makes the curves more difficult to distinguish, resulting in smaller discrepancies and, consequently, reduced power. 

In contrast, the \textit{Kolmogorov-Smirnov} test performs reasonably well, even with 60\% administrative censoring, producing results similar to those obtained with complete data. Thus, when tail information is limited, the \textit{Kolmogorov-Smirnov} statistic, based on the maximum distance between the curves, appears to provide a more reliable measure of the discrepancy between the data and the null hypothesis. This behavior is consistently observed across higher levels of censoring and across all considered scenarios.

\FloatBarrier
\subsection{Random Censored Data}
\label{subsec:Rand_cens}
In the case of purely random censoring, with overall censoring levels comparable to those considered under administrative censoring, the tails are not necessarily as severely truncated. This is illustrated in the right panel of Figure~\ref{fig:fig4}, where, despite a random censoring proportion of $30\%$, the survival function is estimated up to approximately the $94^\mathrm{th}$ percentile.  

\begin{figure}
    \centering
    \includegraphics[width=1\linewidth]{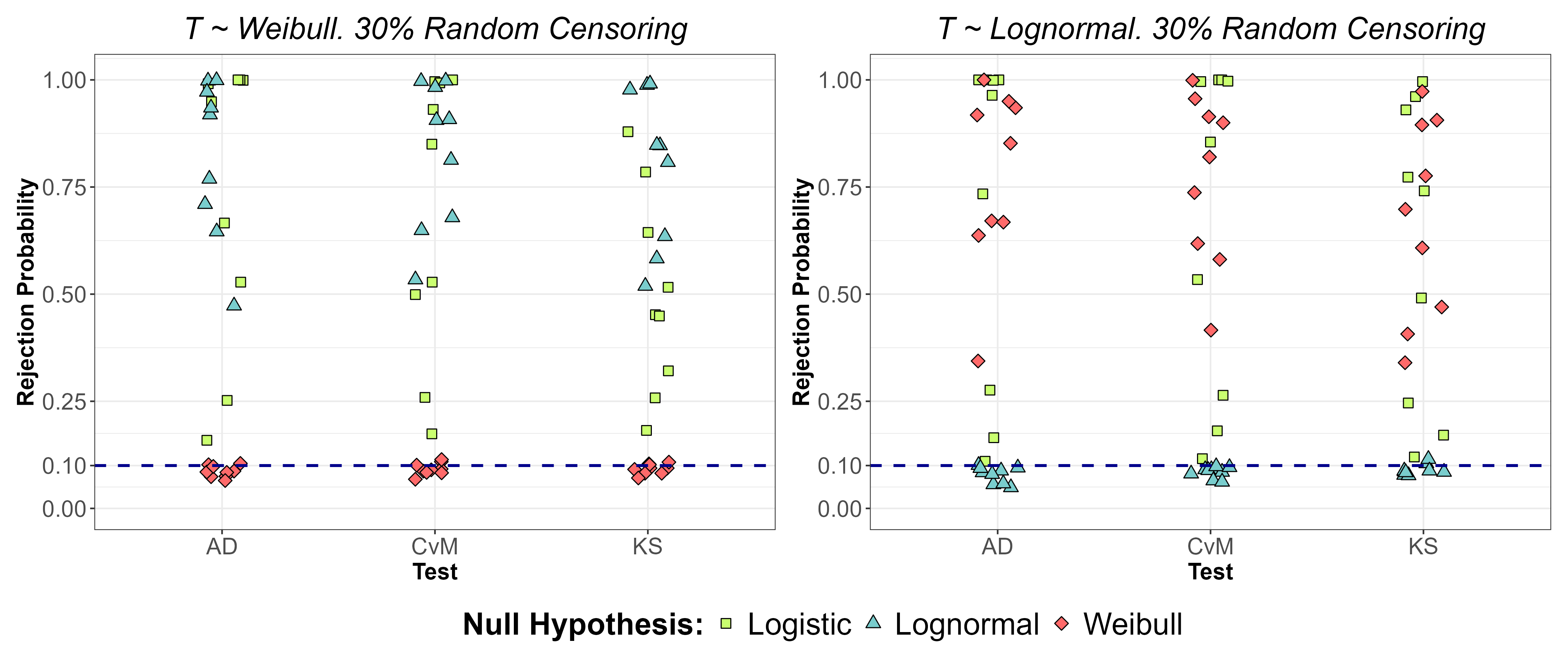}
    \caption{Comparative plots of the estimated rejection probabilities of goodness-of-fit tests (\textit{Anderson-Darling} (AD), \textit{Cramér-von Mises (CvM)}, and \textit{Kolmogorov-Smirnov} (KS), in that order) for 30\% random censored data under different null-hypothesis distributions: logistic (green squares), lognormal (blue triangles), and Weibull (red diamonds). The left panel shows the behavior of the tests for data generated from a Weibull distribution, while the right panel presents the results for data generated from a lognormal distribution. In both panels, for each test and null hypothesis, each point corresponds to a different combination of sample size and degree of skewness. Horizontal jitter has been added to the points to improve the visual separation between overlapping observations.}
    \label{fig:fig6}
\end{figure}

Consequently, as shown in Figure~\ref{fig:fig6}, the power of all tests reaches satisfactory levels, with only a slight reduction compared with the case of complete observations (Figure~\ref{fig:fig3}). This highlights the importance of tail information when conducting goodness-of-fit tests. Moreover, the Type I error is again well controlled at the nominal level. As shown in Figure~\ref{fig:figA2} even with a random censoring proportion of $60\%$, the power is still fairly good in many scenarios.

As with uncensored data, the highest power is observed for the \textit{Anderson-Darling} test and the lowest for the \textit{Kolmogorov-Smirnov} test; however, the differences between the tests are less pronounced than in the case of complete observations. These results suggest that, under random censoring, the relative performance of the tests is less clearly differentiated, as no single method consistently dominates across all scenarios. Thus, the main challenge is to identify which test performs best under each particular scenario.

\begin{figure}
    \centering
    \includegraphics[width=1\linewidth]{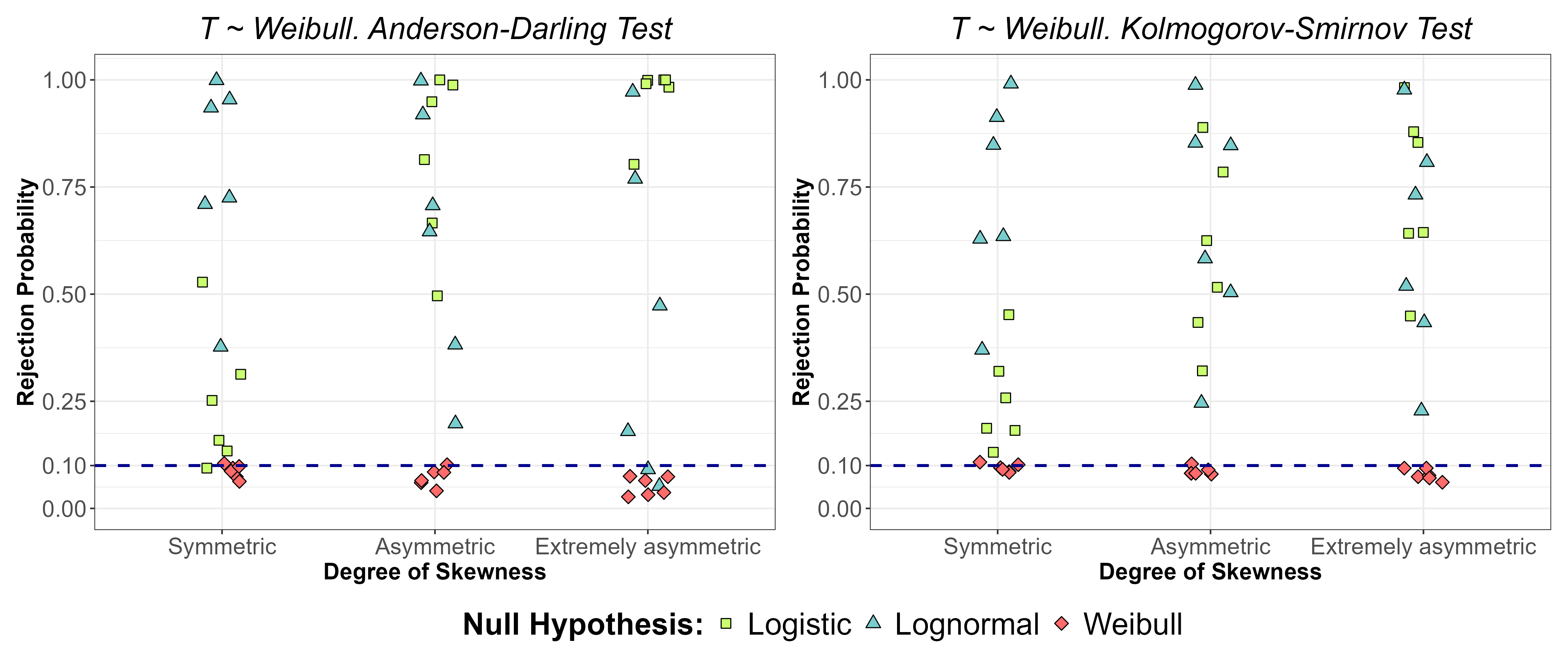}
    \caption{Comparative plots of the estimated rejection probabilities of goodness-of-fit tests (left panel: \textit{Anderson-Darling}; right panel: \textit{Kolmogorov-Smirnov}) for data from Weibull distributions with 30\% and 60\% random censoring and three degrees of skewness under different null-hypothesis distributions: logistic (green squares), lognormal (blue triangles), and Weibull (red diamonds). Each point corresponds to a different combination of sample size and proportion of censored data. Horizontal jitter has been added to the points to improve the visual separation between overlapping observations.}
    \label{fig:fig7}
\end{figure}

Taking a closer look at the role of skewness under random censoring, Figure~\ref{fig:fig7} shows the rejection probabilities of the \textit{Anderson-Darling} and \textit{Kolmogorov-Smirnov} tests as a function of skewness for data generated from Weibull distributions. While there are only minor differences between the tests for symmetric data, the \textit{Anderson-Darling} test shows slightly higher power for highly skewed data. The same observation is made for data from lognormal distributions (Figure~\ref{fig:figA3}). The results of the \textit{Cram\'er-von Mises} test for both distributions (Figure~\ref{fig:figA4}) are very similar to those of the \textit{Anderson-Darling} test.

Concerning sample size, although its effect is not shown explicitly in the figures, it is important to note that, for all tests, power increases as the sample size increases.

Thus, since we have observed results that may appear contradictory and could suggest that the choice of the best-performing test for a given case is somewhat arbitrary, one might conclude that all three tests have similar power under purely random censoring. Nevertheless, the use of the terminal survival discrepancy metric $Diff_{t_{\max}}$ presented in Section~\ref{subsec:Diff} can be helpful in obtaining a specific recommendation.

\begin{figure}
    \centering
    \includegraphics[width=1\linewidth]{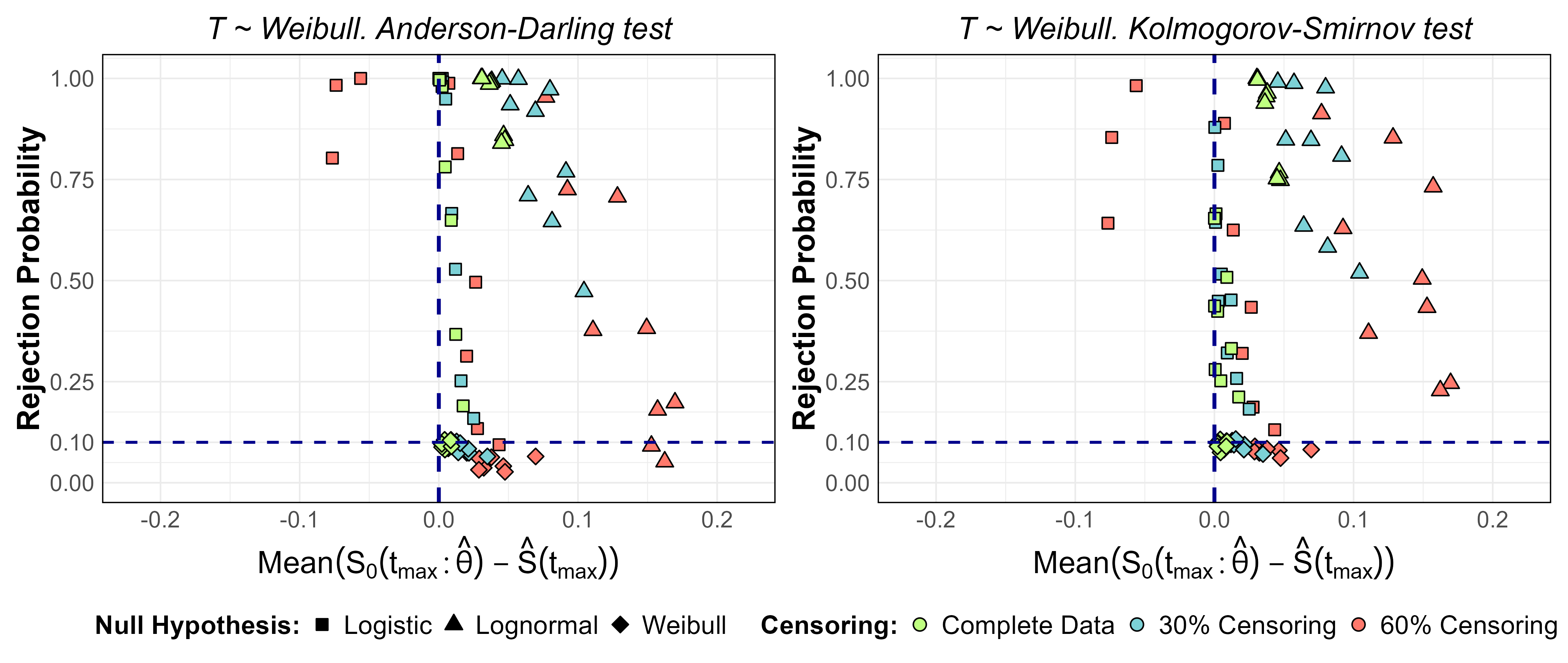}
    \caption{Comparative plots of the estimated rejection probabilities of goodness-of-fit tests (left panel: \textit{Anderson-Darling}; right panel: \textit{Kolmogorov-Smirnov}) for data from Weibull distributions. Rejection probabilities are plotted versus the terminal survival discrepancy metric $Diff_{t_{\max}}$ under different null-hypothesis distributions (logistic: squares; lognormal: triangles; Weibull: diamonds) and degrees of random censoring (Uncensored data: green; 30\% censoring: blue; 60\% censoring: red). Each point corresponds to a different combination of sample size and degree of skewness.}
    \label{fig:fig8} 
\end{figure}

This is illustrated in Figure~\ref{fig:fig8}, which shows the estimated rejection probabilities of the \textit{Anderson-Darling} and \textit{Kolmogorov-Smirnov} tests as a function of $Diff_{t_{\max}}$ for data generated from Weibull distributions. It can be observed that, as the difference becomes more positive, the performance of all three tests deteriorates. One possible driver of this behavior is heavy tail truncation, which occurs either under substantial administrative censoring or when random censoring, by chance, affects the tail of the distribution.

It is also worth noting that the test that declines most gradually under these conditions is the \textit{Kolmogorov-Smirnov} test, which may explain why it was the only test to exhibit acceptable performance in such scenarios (see Section \ref{subsec:Admin_cens}). 

When the difference is zero or close to zero, we observe that the \textit{Kolmogorov-Smirnov} test performs the worst, whereas the \textit{Anderson-Darling} test and \textit{Cramér-von Mises} test (left panel of Figure~\ref{fig:figA6}) have similar behavior, with \textit{Anderson-Darling} performing slightly better. This pattern is consistent with what was observed under complete data conditions (see Section \ref{subsec:Complete_data}), as, when the tails are fully observed, these differences are expected to be minimal or zero. In this case, the maximum likelihood estimation of the null model parameters makes full use of the available information in the data, which may result in better tail fitting.

Finally, in the case of negative differences, which occur more frequently with the data from lognormal distributions (see Figure~\ref{fig:figA5} and the right panel of Figure~\ref{fig:figA6}), the \textit{Anderson-Darling} and \textit{Cramér-von Mises} again perform better, as they remain more stable than the \textit{Kolmogorov-Smirnov} test, which is highly sensitive to sample size and does not outperform the other tests.

Overall, these results highlight the importance of carefully examining tail behavior and the proportion of censoring in order to determine the most appropriate test for each scenario.

Based on these results, a decision-support flow chart and a simplified itemize cheat sheet can be constructed to assist users in selecting the most suitable test for their specific case.

\section{Practical User Guidelines}
\label{sec:Guidelines}
The main objective of this study is to translate our empirical simulation findings into data-driven recommendations for practitioners conducting survival and reliability analyses.

\begin{figure}
    \centering
    \includegraphics[width=0.78\linewidth]{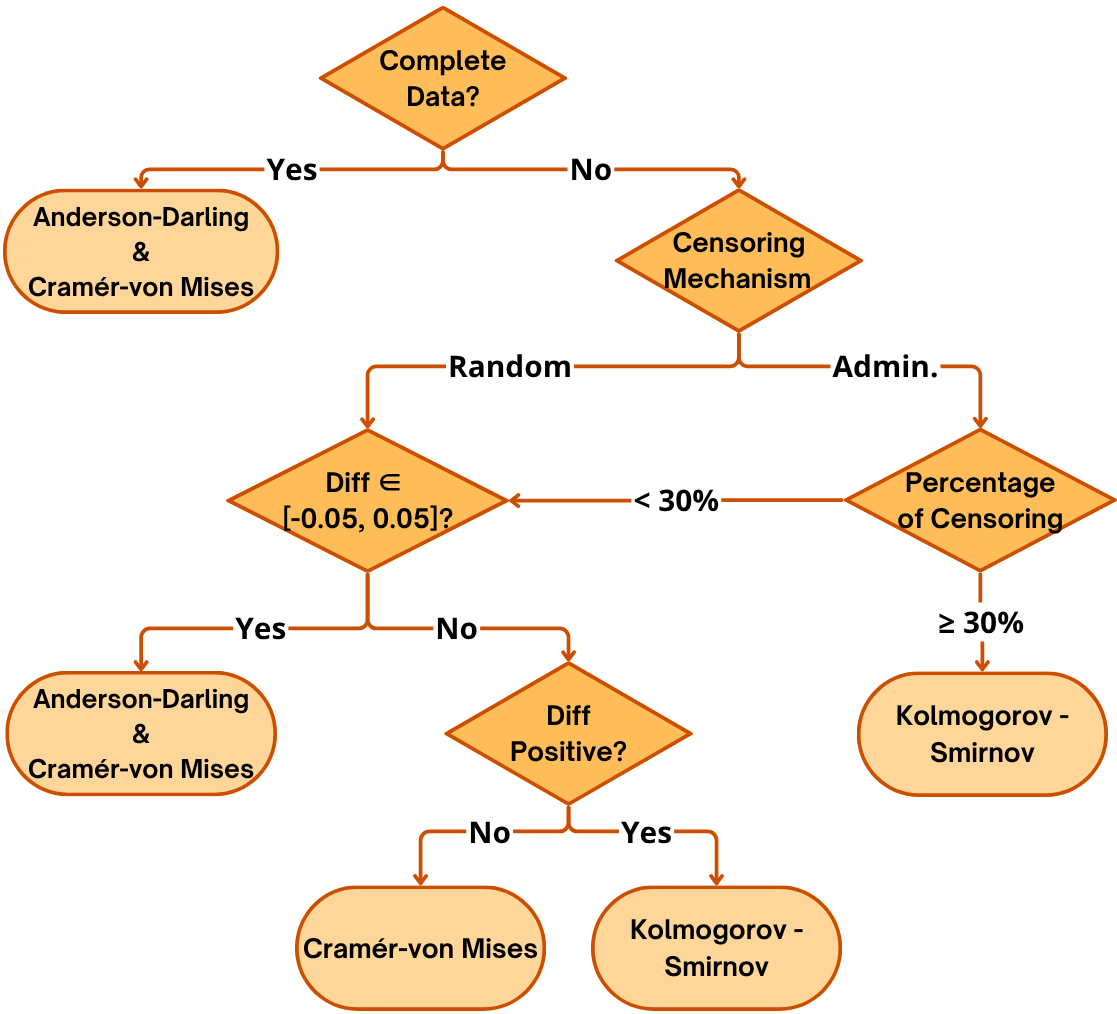}
    \caption{Decision flowchart for selecting a goodness-of-fit test for right-censored survival data. The workflow considers the censoring mechanism, censoring proportion, and the structure of the $Diff_{t_{\max}}$ diagnostic.}
    \label{fig:Decision_Flowchart}
\end{figure}

While administrative censoring generally favors the supremum-based Kolmogorov-Smirnov test in the presence of substantial tail truncation, independent random censoring presents a more complex setting. Under random censoring, several factors, including sample size, distributional skewness, censoring proportion, and upper-tail distortion, affect test performance. Consequently, selecting the most appropriate goodness-of-fit test is less straightforward. To assist researchers in this selection, this section introduces a structured decision protocol centered on the terminal survival discrepancy metric $Diff_{t_{\max}}$, together with the decision flowchart shown in Figure~\ref{fig:Decision_Flowchart}.

In real-world clinical and industrial applications, the underlying data-generating distribution is unknown, and the extent and nature of censoring may not always be fully characterized from the observed data alone. The proposed terminal survival discrepancy metric, $Diff_{t_{\max}}$, provides an easily estimable diagnostic based on the discrepancy between the fitted survival function under a candidate null model and the nonparametric Kaplan-Meier estimate at the largest observed time. Although the metric is defined relative to the candidate null distribution, the simulation results showed consistent patterns across the Weibull, lognormal, and logistic null hypotheses. These patterns can therefore be used as a practical aid for selecting among the goodness-of-fit tests considered in this study.

We propose the following three-step decision framework:
\begin{enumerate}[\bfseries Step 1:]
\item \textbf{Preliminary data characterization} 

  Prior to applying formal goodness-of-fit tests, the user should identify the main characteristics of the observed sample, including:
  \begin{itemize}
    \item \textit{Sample size ($n$):} Determine whether the sample is small ($n \leq 75$), for which test power may be limited and sampling variability may be substantial, or moderate-to-large ($n > 75$).

    \item \textit{Censoring mechanism:} Determine whether censoring is fixed by the study design (administrative) or is associated with random loss to follow-up or other independent censoring mechanisms.

    \item \textit{Censoring proportion:} Assess the proportion of right-censored observations in the sample.
  \end{itemize}

  Distributional skewness may also be assessed descriptively using nonparametric measures such as Bowley's or Kelly's coefficients (Section~\ref{subsec:Skewness}). However, skewness is not used as a separate decision criterion in the proposed framework. Although distributional asymmetry influences test performance, its effect is partly reflected in the terminal survival discrepancy, while the other observed characteristics provide the primary basis for test selection.

\item \textbf{Primary rules of thumb} 

  Evaluate the main structural characteristics of the data:
  \begin{itemize}
    \item \textit{Complete data (uncensored observations):} The area-based \textit{Anderson-Darling} or \textit{Cramér-von Mises} tests are recommended, particularly for moderate-to-large sample sizes, based on their superior empirical power in the simulation study.

    \item \textit{Substantial administrative censoring:} The adapted \textit{Kolmogorov-Smirnov} test is recommended. In the simulation study, the Kolmogorov-Smirnov test was less affected by sharp right-tail truncation than the integrated statistics, whose performance may be reduced by the loss of information beyond the largest observation ($t_{\max}$).
  \end{itemize}

\item \textbf{Random censoring strategy using $Diff_{t_{\max}}$}

  For data subject to random censoring, once a candidate null distribution has been specified and fitted to the observed data, the terminal survival discrepancy can be used as an additional diagnostic for test selection. For a given candidate null distribution, terminal survival discrepancy metric is calculated as in equation~\eqref{eq:diffmetric}.
  
  The simulation results suggest the following empirical rules:
  \begin{itemize}
    \item \textit{Positive discrepancy ($Diff_{t_{\max}}>0.05$)}: The fitted parametric survival function (under the null hypothesis) lies above the Kaplan-Meier estimate at $t_{\max}$. 
    
    In the simulation study, the \textit{Kolmogorov-Smirnov} test generally provided the most robust performance in this setting, whereas the power of the area-based tests was more affected by the resulting tail discrepancy.

    \item \textit{Near-zero discrepancy ($|Diff_{t_{\max}}| \leq 0.05$)}: The fitted parametric survival function and the nonparametric survival curves are closely aligned at $t_{\max}$. 
    
    In this setting, the area-based tests generally performed best, with the \textit{Anderson-Darling} test achieving the highest overall empirical power, followed by the \textit{Cramér-von Mises} test.

    \item \textit{Negative discrepancy ($Diff_{t_{\max}}<-0.05$)}: The fitted parametric survival function lies below the Kaplan-Meier estimate at $t_{\max}$. 
    
    In the simulation study, the \textit{Cramér-von Mises} test generally exhibited the highest power, followed by the \textit{Anderson-Darling} test.
  \end{itemize}
The threshold of $0.05$ is empirically motivated by the simulation results and should not be interpreted as a universal cutoff.
\end{enumerate}

The full selection process is summarized in the decision flowchart presented in Figure~\ref{fig:Decision_Flowchart}.

The recommendations presented above are necessarily subject to the scope of the simulation study. In particular, they are based on a finite set of simulation settings, including three sample sizes ($n=75$, $150$, and $300$), three censoring proportions ($0\%$, $30\%$, and $60\%$), three levels of skewness, two data-generating distributions, and two censoring mechanisms. Consequently, the proposed rules and the empirical threshold for $Diff_{t_{\max}}$ should not be interpreted as universal guarantees of test performance. Rather, they provide empirically motivated guidance for data settings comparable to those investigated in this study.

In summary, no single goodness-of-fit test consistently performs best for right-censored survival data across all settings. By combining empirical simulation evidence with the $Diff_{t_{\max}}$ diagnostic, the proposed framework provides a structured approach to selecting a goodness-of-fit test based on characteristics observable in the data. The protocol is intended as a decision-support tool rather than an automated decision rule, and its recommendations should therefore be considered alongside domain-specific knowledge and considerations related to the study design and follow-up process.

As mentioned previously, all adapted tests described in this study are implemented in the R package \texttt{GofCens} \citep{gofcensPackage, gofcensjournal}.
\FloatBarrier
\section{Discussion}
\label{sec:Discussion}
This simulation study provides a systematic evaluation of the performance of classical goodness-of-fit tests under different right-censoring mechanisms and distributional scenarios. Overall, our results highlight that censoring does not affect all goodness-of-fit tests equally. While integrated distance measures such as the \textit{Cramér-von Mises} and \textit{Anderson-Darling} tests generally provide higher power under complete data or mild random censoring, their performance can deteriorate substantially when the upper tail of the distribution is partially unobserved. Conversely, the \textit{Kolmogorov-Smirnov} test, although often less powerful under ideal conditions, showed greater robustness in scenarios with severe loss of information.

Under complete data and low levels of random censoring, the \textit{Anderson-Darling} test consistently achieved the highest empirical power among the evaluated methods. This behavior is consistent with its weighted formulation, which assigns greater importance to deviations occurring in the tails of the distribution. This characteristic is particularly advantageous when the alternative distribution differs from the null hypothesis mainly in extreme regions. Similarly, the \textit{Cramér-von Mises} test showed competitive performance, as it accumulates discrepancies across the entire distribution rather than focusing on a single point of maximum deviation.

However, the same sensitivity to tail deviations that benefits the \textit{Anderson-Darling} test under complete data becomes a limitation under heavy administrative censoring. When administrative censoring is introduced by fixing the end of the follow-up period, subjects who have not experienced the event by the end of the study remain censored, resulting in less information about the upper tail of the survival distribution. In this situation, the Kaplan-Meier estimator may exhibit a large discrepancy with respect to the theoretical distribution near the maximum observed time, artificially increasing the integrated distances used by the \textit{Anderson-Darling} and \textit{Cramér-von Mises} tests. Consequently, both tests showed important deviations from the nominal Type I error rate under strong administrative censoring, indicating that their calibration may be compromised when the observation window does not adequately capture the distributional tail.

In contrast, the \textit{Kolmogorov-Smirnov} test maintained better Type I error control under these challenging scenarios. This robustness is related to its supremum-based definition, which depends only on the maximum vertical distance between the estimated and theoretical distribution functions, avoiding the accumulation of discrepancies over regions with limited information. The analysis of the $Diff_{t_{\max}}$ metric supports this interpretation, as larger discrepancies between the Kaplan-Meier estimate and the theoretical survival curve at the last observed time were associated with a deterioration in the performance of integrated statistics. Therefore, the degree of right-tail distortion provides useful information when selecting an appropriate goodness-of-fit test.

Unlike administrative censoring, random censoring distributes the loss of information throughout the follow-up period, preserving part of the distributional structure and allowing the evaluated tests to maintain more stable behavior. These findings indicate that the impact of censoring is not determined exclusively by the percentage of censored observations, but rather by how the censoring mechanism modifies the observable region of the survival distribution.

Furthermore, the shape of the underlying distribution influenced the relative performance of the tests. Symmetric distributions generally produced smaller differences between the evaluated methods, whereas highly skewed distributions, which are common in survival applications, increased the relevance of selecting an appropriate goodness-of-fit procedure. In particular, scenarios with strong positive skewness showed that the \textit{Kolmogorov-Smirnov} test generally performed better when the data were substantially affected by censoring. These findings indicate that distributional shape can influence the relative performance of the goodness-of-fit tests and should be considered when interpreting their behavior under different censoring conditions.

It is important to note that not all factors investigated in the simulation study are directly incorporated into the proposed practical guidelines. In particular, the underlying distribution and degree of skewness cannot generally be known with certainty from the observed, potentially censored data and therefore are not used as direct decision criteria. Nevertheless, examining test performance across different distributions and levels of skewness is important for assessing the robustness of the methods and for identifying patterns that can be translated into diagnostics based on observable characteristics of the data, such as $Diff_{t_{\max}}$.

Despite the extensive simulation design, several limitations should be acknowledged. As with any simulation study, it is not possible to consider all possible combinations of distributions, parameter values, sample sizes, censoring mechanisms, and other features of survival data. The conclusions are therefore necessarily restricted to the settings investigated here. In particular, the study considered three sample sizes ($n=75$, $150$, and $300$), three censoring proportions ($0\%$, $30\%$, and $60\%$), three levels of skewness, two data-generating distributions, and two censoring mechanisms.

Although the Weibull and lognormal distributions used to generate the data, together with the Weibull, lognormal, and logistic null distributions considered in the goodness-of-fit tests, cover several common survival scenarios, other distributional families with different hazard structures, such as gamma, Gompertz, or log-logistic distributions, may exhibit different behavior. Similarly, the study focused on symmetric and right-skewed distributions, while left-skewed scenarios were not investigated. Such situations may be particularly relevant under administrative censoring, as the most informative part of the distribution could become hidden beyond the observation window. Finally, random censoring was assumed to be independent of the event time. Future studies should extend this framework to a broader range of distributions and parameter configurations, larger and smaller sample sizes, and also consider informative censoring. Such extensions would help assess the generalizability and robustness of the proposed recommendations.
\FloatBarrier

\section*{Acknowledgments}
This research was funded by MICIU/AEI/10.13039/2501100011033 and FEDER, UE [PID2023-148033OB-C21] and by Agència de Gestió d’Ajuts Universitaris i de Recerca [2021 SGR 01421] (GRBIO) from Generalitat de Catalunya (Spain).

We thank Mireia Besal\'u, Matilde Francisco, and Arnau Garcia-Fern\'andez for their valuable contributions to the early stages of this work.

\bibliographystyle{cas-model2-names}

\bibliography{PaperFLBG}

\FloatBarrier
\appendix
\setcounter{figure}{0}
\renewcommand{\thefigure}{\thesection\arabic{figure}}

\clearpage
\section{Simulation Study Results: Further Figures}
\label{sec:appA}
\begin{figure}[h]
\centering
  \includegraphics[width=1\linewidth]{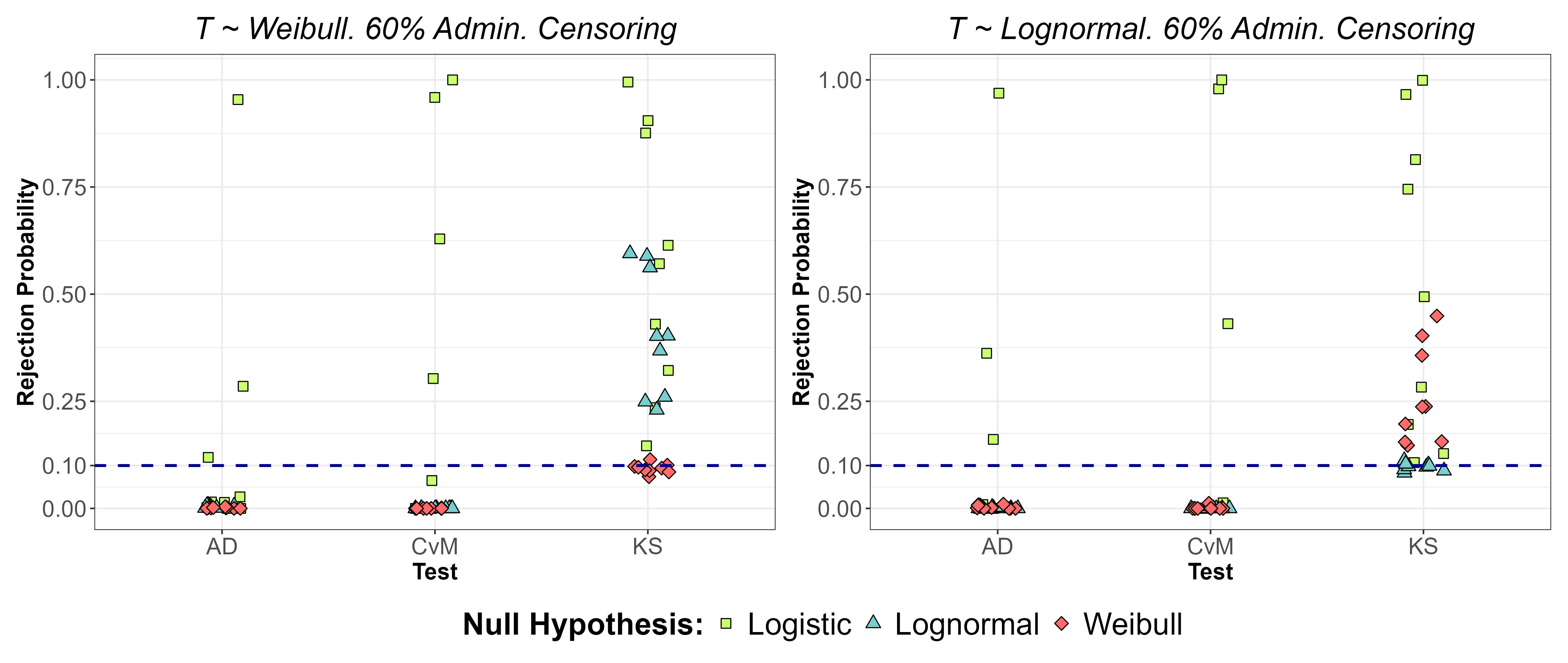}
  \caption{Comparative plots of the estimated rejection probabilities of goodness-of-fit tests (\textit{Anderson-Darling} (AD), \textit{Cramér-von Mises (CvM)}, and \textit{Kolmogorov-Smirnov} (KS), in that order) for 60\% administrative censored data under different null-hypothesis distributions: logistic (green squares), lognormal (blue triangles), and Weibull (red diamonds). The left panel shows the behavior of the tests for data generated from a Weibull distribution, while the right panel presents the results for data generated from a lognormal distribution. In both panels, for each test and null hypothesis, each point corresponds to a different combination of sample size and degree of skewness. Horizontal jitter has been added to the points to improve the visual separation between overlapping observations.}
  \label{fig:figA1}
\end{figure}

\begin{figure}[h]
    \centering
    \includegraphics[width=1\linewidth]{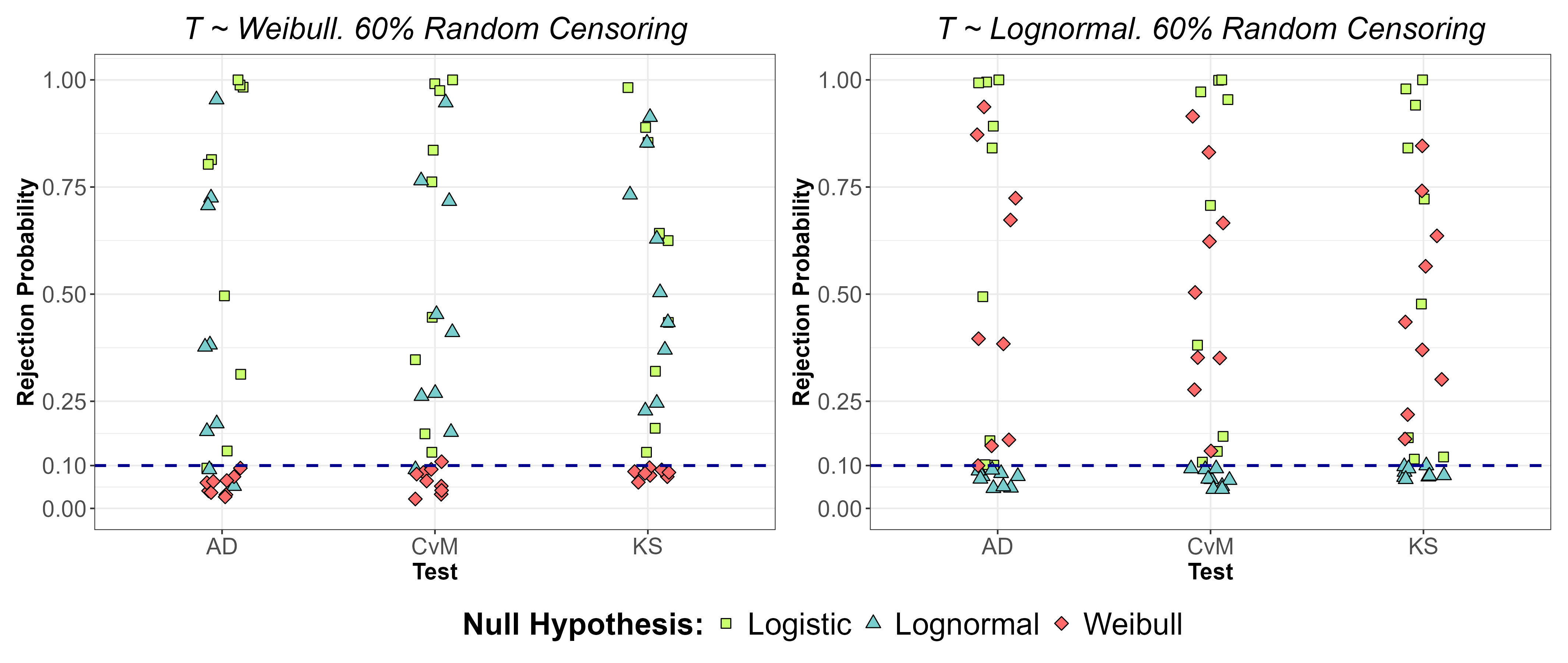}
    \caption{Comparative plots of the estimated rejection probabilities of goodness-of-fit tests (\textit{Anderson-Darling} (AD), \textit{Cramér-von Mises (CvM)}, and \textit{Kolmogorov-Smirnov} (KS), in that order) for 60\% random censored data under different null-hypothesis distributions: logistic (green squares), lognormal (blue triangles), and Weibull (red diamonds). The left panel shows the behavior of the tests for data generated from a Weibull distribution, while the right panel presents the results for data generated from a lognormal distribution. In both panels, for each test and null hypothesis, each point corresponds to a different combination of sample size and degree of skewness. Horizontal jitter has been added to the points to improve the visual separation between overlapping observations.}
    \label{fig:figA2}
\end{figure}

\begin{figure}[h]
    \centering
    \includegraphics[width=1\linewidth]{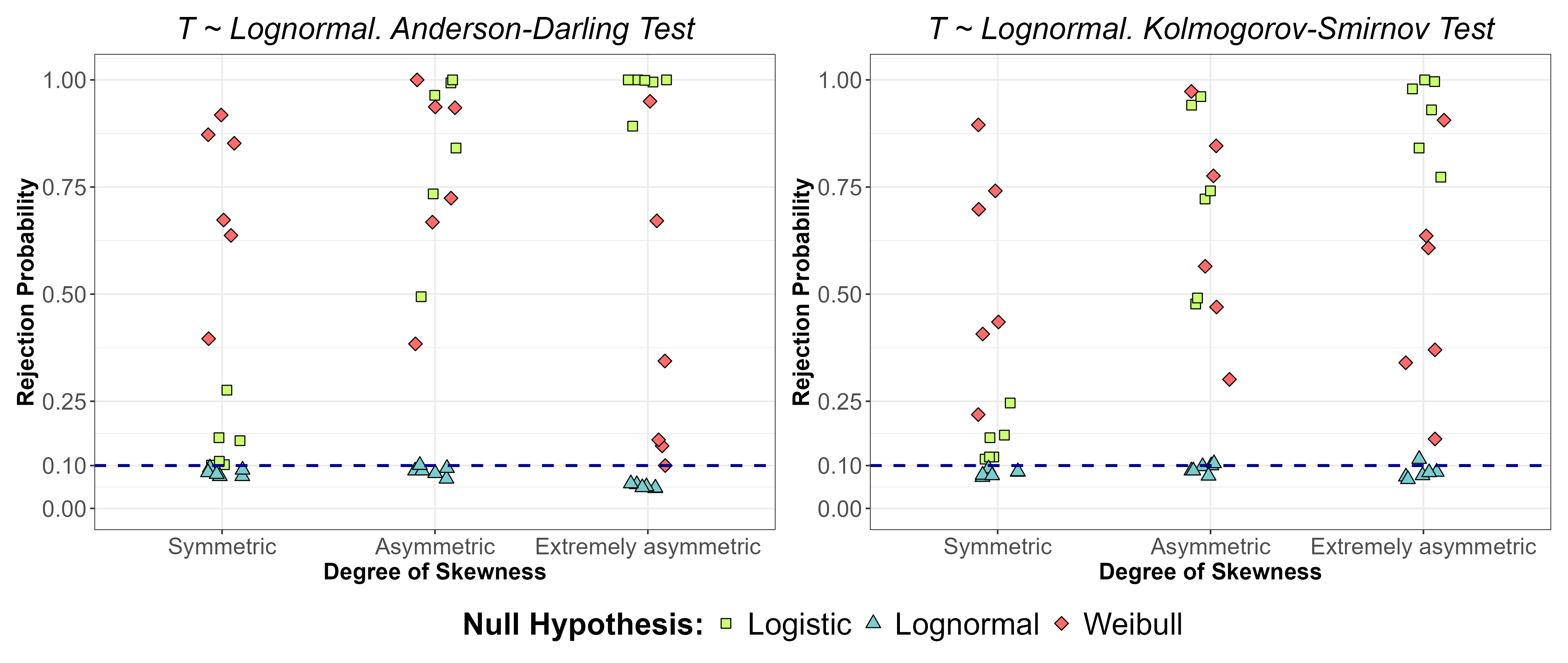}
    \caption{Comparative plots of the estimated rejection probabilities of goodness-of-fit tests (left panel: \textit{Anderson-Darling}; right panel: \textit{Kolmogorov-Smirnov}) for data from lognormal distributions with 30\% and 60\% random censoring and three degrees of skewness under different null-hypothesis distributions: logistic (green squares), lognormal (blue triangles), and Weibull (red diamonds). Each point corresponds to a different combination of sample size and proportion of censored data. Horizontal jitter has been added to the points to improve the visual separation between overlapping observations.}
    \label{fig:figA3}
\end{figure}

\begin{figure}[h]
    \centering
    \includegraphics[width=1\linewidth]{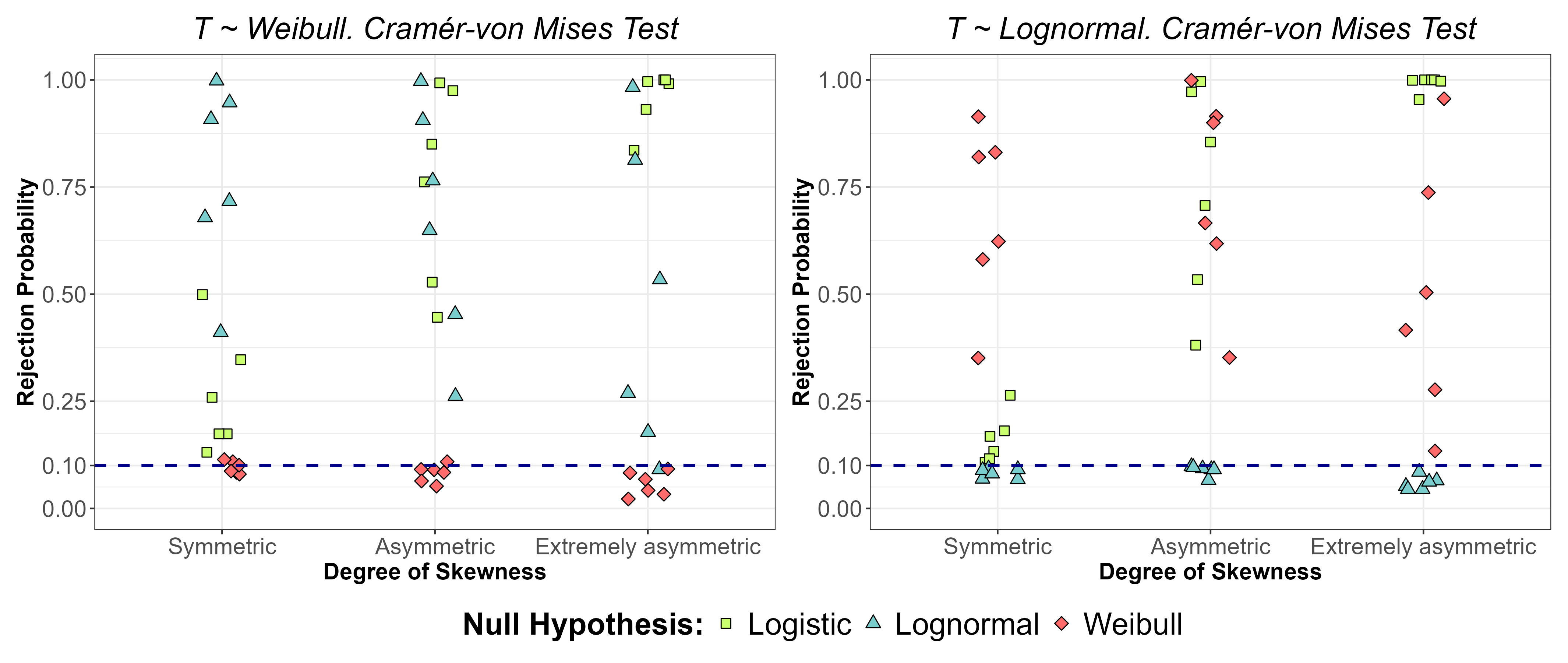}
    \caption{Estimated rejection probabilities of \textit{Cram\'er-von Mises} test for data from Weibull (left panel) and lognormal (right panel) distributions with 30\% and 60\% random censoring and three degrees of skewness under different null-hypothesis distributions: logistic (green squares), lognormal (blue triangles), and Weibull (red diamonds). Each point corresponds to a different combination of sample size and proportion of censored data. Horizontal jitter has been added to the points to improve the visual separation between overlapping observations.}
    \label{fig:figA4}
\end{figure}

\begin{figure}[h]
    \centering
    \includegraphics[width=1\linewidth]{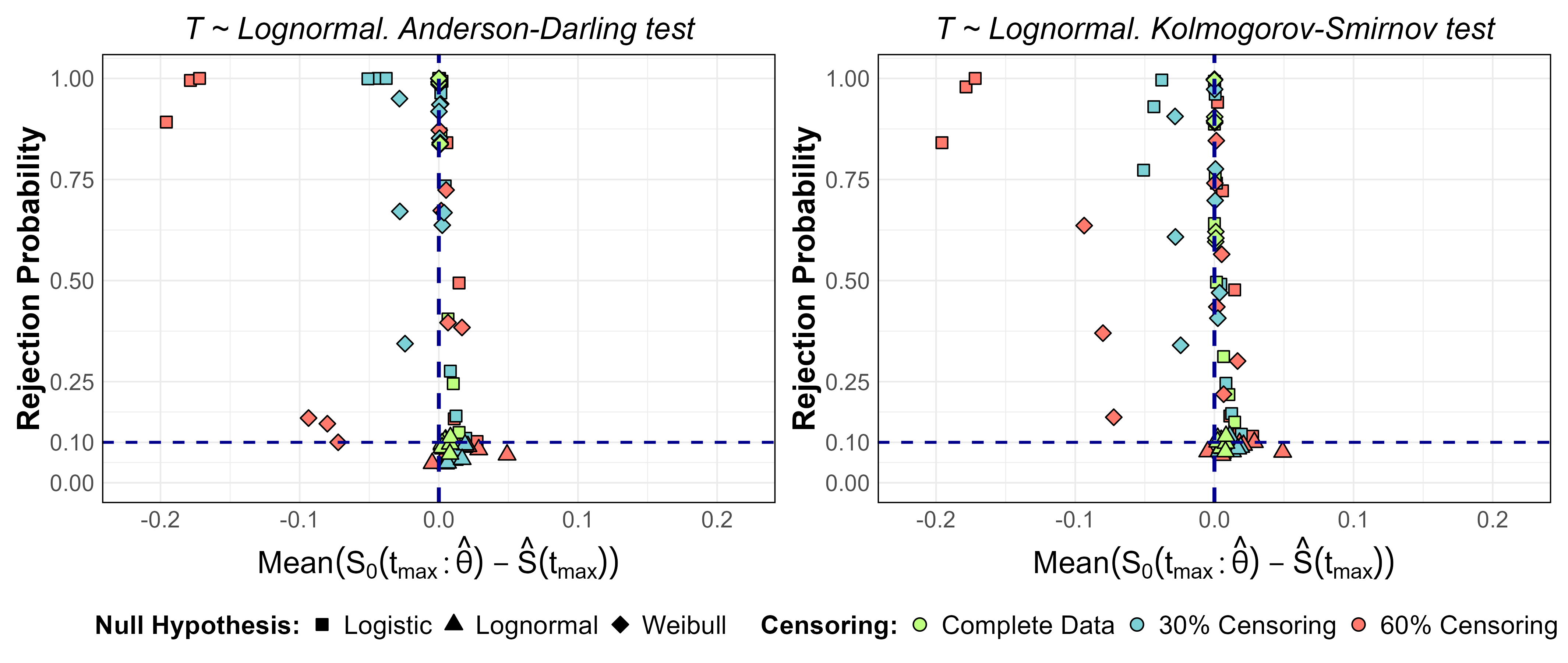}
    \caption{Comparative plots of the estimated rejection probabilities of goodness-of-fit tests (left panel: \textit{Anderson-Darling}; right panel: \textit{Kolmogorov-Smirnov}) for data from lognormal distributions. Rejection probabilities are plotted versus the terminal survival discrepancy metric $Diff_{t_{\max}}$ under different null-hypothesis distributions (logistic: squares; lognormal: triangles; Weibull: diamonds) and degrees of random censoring (Uncensored data: green; 30\% censoring: blue; 60\% censoring: red). Each point corresponds to a different combination of sample size and degree of skewness.}
    \label{fig:figA5}
\end{figure}

\begin{figure}[h]
    \centering
    \includegraphics[width=1\linewidth]{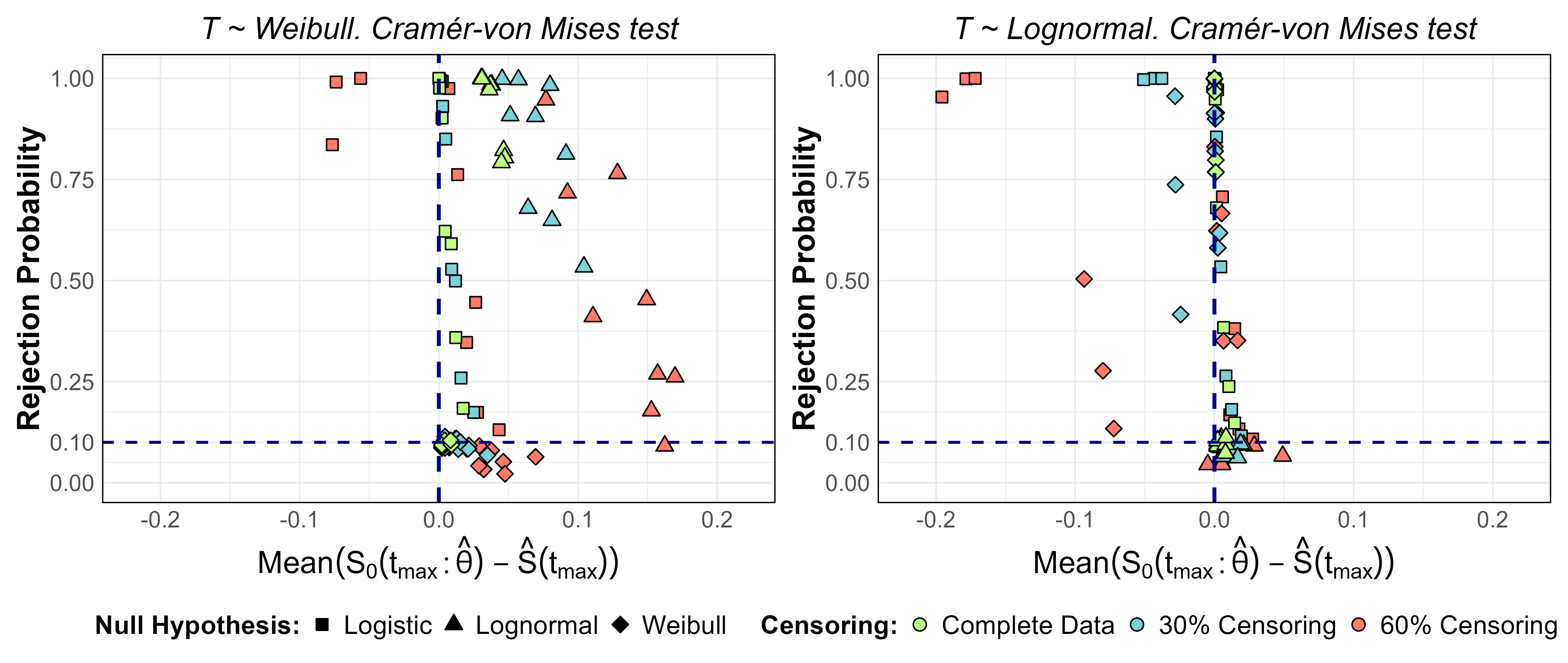}
    \caption{Estimated rejection probabilities of \textit{Cram\'er-von Mises} test for data from Weibull (left panel) and lognormal (right panel) distributions. Rejection probabilities are plotted versus the terminal survival discrepancy metric $Diff_{t_{\max}}$ under different null-hypothesis distributions (logistic: squares; lognormal: triangles; Weibull: diamonds) and degrees of random censoring (Uncensored data: green; 30\% censoring: blue; 60\% censoring: red). Each point corresponds to a different combination of sample size and degree of skewness.}
    \label{fig:figA6}
\end{figure} 
\end{document}